\documentclass[12pt]{elsarticle}
\usepackage{amsmath,amssymb,amsfonts}
\usepackage{mathrsfs,mathptmx,mathtools}
\usepackage{color}
\usepackage{float}
\usepackage{graphicx}
\usepackage{mwe}
\usepackage{subfigure}
\usepackage{tikz}
\usepackage{epstopdf}
\usepackage{pgfplots}
\usetikzlibrary{arrows}
\usepackage{lineno}
\usepackage{changes}
\usepackage{hyperref}
\usepackage{url}
\usepackage{epsfig}
\usepackage[margin=2.5cm]{geometry}
\hypersetup{
    colorlinks=true,
    linkcolor=blue,
}
\usepackage{multirow}
\usepackage{bm}
\usepackage{amsthm}
\usepackage{tikz}

\usepackage{caption}
\usepackage{subcaption}

\usepackage{color}
\usepackage{pgfplots}
\usepackage[symbol]{footmisc}
\tikzstyle{nicebox}=[draw=black!100, fill=white!10, rectangle, inner sep=4pt, inner ysep=16pt]
\tikzstyle{niceboxtitle}=[draw=black!100, fill=white, text=black, rectangle]

\usetikzlibrary{arrows,decorations.markings}
\usetikzlibrary{plotmarks}
\biboptions{sort&compress}

\usepackage[normalem]{ulem}

\usepackage{cleveref}
\usepackage{citeref}

\biboptions{sort&compress}

\theoremstyle{remark}

\newcommand{\xx}{\mathbf{x}}

\newcommand{\bfsn}{\mathbf{n}}
\newcommand{\bfn}{\mathbf{N}}

\newcommand{\bl}{\boldsymbol{\lambda}}
\newcommand{\bpsi}{\boldsymbol{\Psi}}

\graphicspath{{./Figures/}}

\usepackage{cleveref}

\begin{document}

\begin{frontmatter}
\title{Sloshing reduction in a swaying tank with porous baffles using scaled boundary finite element method}

\author[iitm]{Pramod ALN}

\author[iitm2]{Aditi Choudhury}
\author[iitm2]{KG Vijay}
\author[ballarat]{Ean Tat Ooi}
\author[iitm]{Sundararajan Natarajan\corref{cor}\fnref{fnlab1}}
\address[iitm]{Department of Mechanical Engineering, Indian Institute of Technology Madras, Chennai - 600036, Tamil Nadu, India.}
\address[iitm2]{Department of Ocean Engineering, Indian Institute of Technology Madras, Chennai - 600036, Tamil Nadu, India.}
\address[ballarat]{School of Science, Engineering \& Information Technology, Federation University Australia, Ballarat VIC3350, Australia.}

\cortext[cor]{Corresponding author}
\fntext[fnlab1]{Department of Mechanical Engineering, Indian Institute of Technology Madras, Chennai - 600036, Tamil Nadu, India. Email: snatarajan@iitm.ac.in}

\begin{abstract}

Sloshing is an inevitable phenomenon in an ocean-going vessel that can have adverse effects. In this work, the mitigation of sloshing is investigated using multiple thin porous baffles of various configurations in a partially filled swaying tank. The boundary value problem is solved within the framework of a linearized potential flow theory using the scaled boundary finite element method (SBFEM). The flow through the thin porous baffles is assumed to follow Darcy's law. The computational domain is divided into a minimum number of subdomains due to the presence of porous baffles and to ensure star convexity. Higher-order polynomials are used along each subdomain edge to represent the unknown field, i.e., velocity potential. The developed numerical model is validated with the known results in the literature. Subsequently, various results, such as the amplification factor and the forces of the tank wall, are presented and discussed for the effect of configuration, porosity, slosh tank width, depth of baffle submergence and the space between adjacent baffles. From the parametric study, it is observed that top-mounted baffles enhance sloshing suppression by $50\%$ compared to bottom-mounted vertical baffles, considering all sloshing modes. Assessing the overall effectiveness, top-mounted convex baffle configuration emerges as the most efficient configuration for sloshing suppression, achieving a well-balanced reduction across all modes.  
\end{abstract}


\end{frontmatter}

\section{Introduction}
Sloshing refers to the undesirable motion of a liquid's free surface in a partially filled container caused by external forces. This is widely prevalent in ocean-going vessels~\cite{LI2019246,pujarirajan2024,yuwu2025}, aircraft, rockets~\cite{cui2014parametric}, seismic storage containers~\cite{merino2020probabilistic,zhangchen2024}, and nuclear vessels~\cite{MYRILLAS2017317}. Due to continuous excitation, the relative motion of the liquid's free surface exerts loading on the tank side walls. This is remarkably high in tanks having a very high aspect ratio (tank width/fill depth). This is termed the free surface effect, which leads to overturning or structural damage, potentially resulting in catastrophic failure.

The excessive movement of liquid in containers can be reduced by either compartmentalization or suppressors such as partially submerged impermeable or porous baffles are employed. Porous baffles are preferred to impermeable baffles because they enhance damping~\cite{tang2024effect} and suppress the dynamic forces that act on the tank~\cite{cho2008wave}. Various configurations of porous baffles (horizontal baffles~\cite{jin2014experimental}, vertical baffles~\cite{cho2016effect}, wall-mounted ring baffle~\cite{ZANG2019101963}, T-shaped baffle~\cite{KARGBO2021108664}) are used to vary the sloshing frequency and amplitude of the liquid in the tank. The free surface elevation can be sensibly tuned by the appropriate selection of the structural parameters such as the tank width, baffle configuration, baffle porosity~\cite{gao2021finite} and also the length of the baffle~\cite{biswal2004dynamic}. \\

Analytical solutions for various configuration of tanks were developed~\cite{FALTINSEN_TIMOKHA_2010,ibrahim2005liquid} based on linear potential theory to characterize the natural frequency and to reduce the sloshing phenomena. 
The restriction of analytical solutions to simple geometries and limited baffle configurations lead to the development of finite element method (FEM)~\cite{cho2005finite} based models. Further FEM has been used to understand the sloshing behaviour of tanks with cylindrical~\cite{mitra2008slosh}, rectangular~\cite{kumar2014dynamics,kumar2016dynamics,gao2021finite}, prismatic~\cite{Kumar_2021} geometries. Alternatively, the boundary element method (BEM) discretizes only the boundaries, effectively reducing the problem's spatial dimension by one.  
BEM has been employed to study sloshing dynamics in both rectangular and cylindrical tanks~\cite{firouz20083d, sygulski2011boundary, hu2018natural, guan2020numerical} and baffles with various configurations~\cite{zang2019boundary}. However, BEM requires the availability of a fundamental solution for modeling the behavior at the boundaries of the domain. Obtaining or deriving this solution can be mathematically and computationally challenging due to the variability with geometric and boundary conditions, as well as the problem-specific complexity of the solution. \\

The scaled boundary finite element method (SBFEM) introduced by Wolf and Song~\cite{song2000scaled}.
The semi-analytical approach of the SBFEM solves the partial differential equation in the circumferential direction numerically and the resulting ordinary differential equation analytically in the radial direction. Since its inception, the SBFEM proved to be efficient for dynamics~\cite{SONG1997329,struc_dynamics}, fracture mechanics~\cite{SONG201845,SONG2002183,CHIONG2014210}, acoustics~\cite{KHAJAH2021102732,BIRK2016252}, contact mechanics~\cite{XING2018114,XING2019928,HIRSHIKESH2021104180} and piezoelectric~\cite{LI201352,OOI2015101} problems. Another attractive feature of the SBFEM is that the method allows to choose elements of $n-$sides and polynomial of varying orders~\cite{higherorder_SBFEM} can be used along each edge of an element. Teng et al.~\cite{teng2006} and Lin et al.~\cite{lin2015scaled} used SBFEM to understand sloshing of liquid in baffle-free tank with linearized free surface boundary condition. Wang et al.~\cite{Wang2016_ell_cy, WANG2016_cy_multibaff, Wang2017_t_shaped, WANG2017_tor_tank} did a series of works on 2D and 3D tanks of different shapes with rigid baffles. In~\cite{YE2024_fluid_str} Wenbin Ye et al. considered an elastic tank wall by taking tank-fluid coupling and studied liquid sloshing in baffle free rectangular tank. In all these cases, SBFEM has been used for baffle-free or impermeable baffles.
The application of SBFEM used to reduce the amplification factor and sloshing force on the wall in cylindrical tanks with porous structures is that of Wenbin~\cite{ye2018application}. Recently, Zang et al.,~\cite{zangzhang2024} employed SBFEM to study sloshing in rectangular tanks, however, their study was limited to rigid baffles.\\



To the best of the author's knowledge, investigations on the sloshing dynamics using SBFEM are scanty. This work proposes to use the SBFEM to understand the sloshing of liquids in tanks with multiple porous baffles of various configurations. The pressure difference across the thin, rigid porous baffle is assumed to be linearly proportional to the flow velocity (Darcy's law). Numerical results are obtained by solving the boundary value problem within the linear potential flow theory framework. The computational domain is discretized into subdomains based on the porous baffle configurations in the tank. Along each edge of the subdomain, higher-order polynomials are used. Numerical examples showed that the present technique predicts the sloshing in liquid tanks with less computational effort without compromising in accuracy.  \\


The rest of the paper is organized as follows: the governing equation and the boundary conditions describing the sloshing in two dimensional tanks with and without baffle is described in \Cref{sec:goveqn}. \Cref{sec:sbfemintro} presents the scaled boundary finite element method for numerically studying the sloshing in rectangular tanks. The efficacy, accuracy and the convergence properties are studied in detail in \Cref{sec:numexam}, followed by major conclusions in the last section.

\section{Mathematical Formulation}
\label{sec:goveqn}
\begin{figure}[htpb]
    \centering
    \includegraphics[scale=0.25]{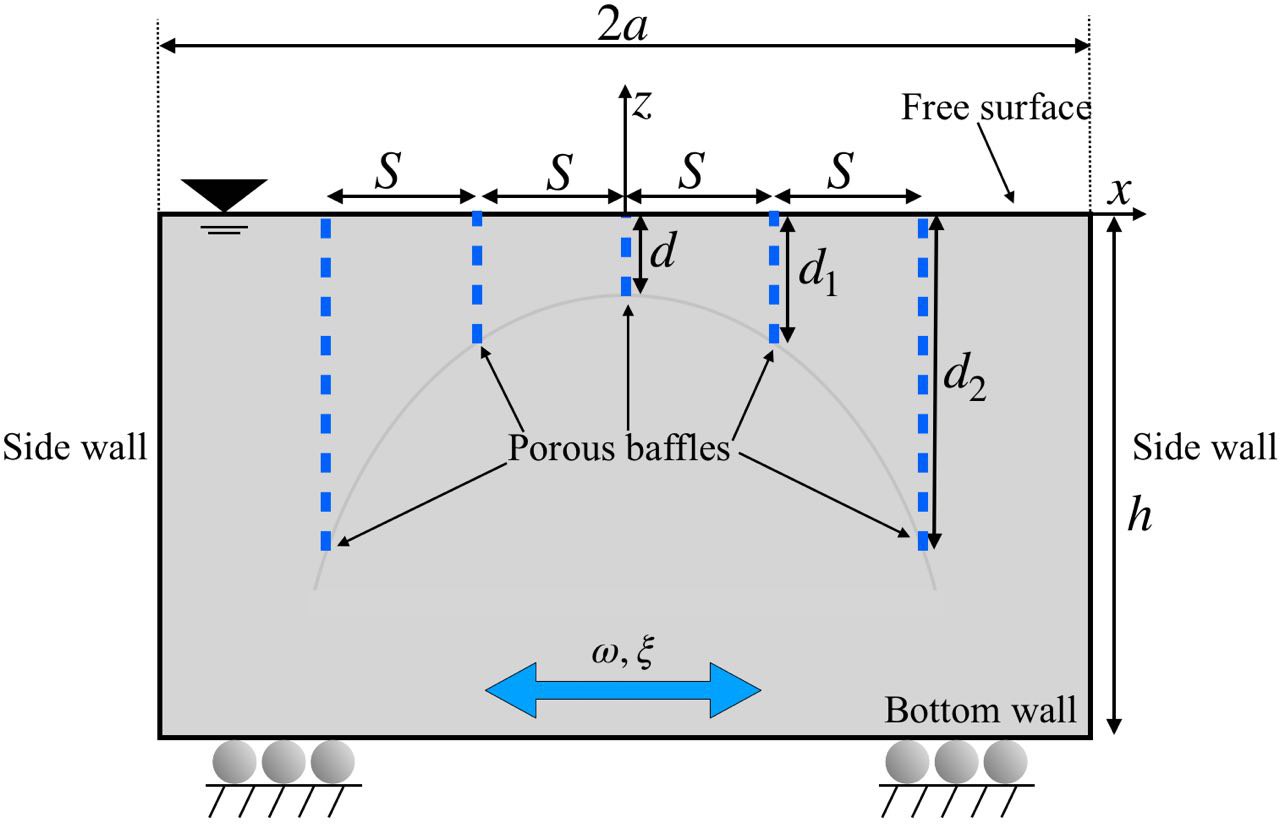}
    \caption{Schematic representation of a rectangular tank with five top-mounted vertical baffle where the endpoints of the baffles follow a parabolic profile}
    \label{case-A}
\end{figure}

In this section, we investigate the two-dimensional sloshing phenomena in a rectangular tank of the dimension $2a \times h$ with multiple vertical porous baffles. A Cartesian coordinate system is adopted, with the $x$-axis oriented horizontally and the $z$-axis directed vertically upward as mentioned in~\Cref{case-A}. The endpoints of the baffles follow a parabolic profile. 
The heights of the middle and extreme side baffle are denoted by $d$ and $d_2$, respectively. Depending on the choice of the parabolic profile (convex or concave), the center baffle will be the shortest or longest, and vice-versa in the case of extreme-side baffles. For instance, in \Cref{case-A}, the middle baffle has the smallest height $(d)$ whereas the extreme-side baffles has the largest height $(d_2)$. 
The general equation of parabola is,
\begin{equation}
    x^2 = 4\alpha (y+y_0)
    \label{parabola_eq}
\end{equation}
where, in our case  $y_0>0$ for all parabolic configurations. Each baffle has a free tip that is in contact with the liquid inside the tank. Given the height of the middle and extreme-side baffles ($d$ and $d_2$),the distance of the free tip from the free surface, as well as the coordinates of the free tip, can be determined.  The value of $y_0$ in~\Cref{parabola_eq} is obtained from the distance between the free surface and the free tip of the middle baffle. Once $y_0$ is determined, the only unknown in the equation of the parabola is $\alpha$. The coordinates of the free tip of the extreme baffle must satisfy the parabolic equation, allowing us to solve for $\alpha$. Once the equation of the parabola is established, the height (or $y$-coordinate) of an intermediate baffle can be determined for a given $x$-coordinate, denoted as $S$.  Specifically, for \Cref{case-A}, we have $y_0 = d$, and the coordinate of the last baffle is $(2S,-d_2)$. This leads to the relation $\alpha = -S^2/(d_2-d)$. The height of the middle baffle is then given by $d_1 = |-0.25(d_2-d)-d| = 0.25(d_2-d)+d$. A similar approach can be applied for other baffle configuration where the baffle end points follow a parabolic profile.
The tank undergoes forced horizontal oscillations with an amplitude $\xi$ and frequency $\omega$. The fluid is assumed to be inviscid, incompressible and irrotational. Since the wave motions are small, linear potential theory is applicable. The fluid particle velocity is expressed as the gradient of the velocity potential $\Phi(x,z,t)$  which, under the assumption of harmonic motion with angular frequency $\omega$, takes the form $\Phi(x,z,t)= Re\{-i\omega \xi \phi(x,z) e^{-i\omega t}\}$. \\

The governing equations for two dimensional sloshing response is given by:
\begin{equation}
    \label{eqn:governeqn}
    \nabla \cdot \nabla \phi = 0, \qquad {\rm in} \quad \Omega
\end{equation}
supported by the following boundary conditions on the external boundary:
\begin{equation}
\left\{
\begin{aligned}
    \nabla \phi \cdot \bfsn &= \dfrac{\omega^2}{g} \phi \qquad {\rm on} \quad z = 0 \\
    \nabla \phi \cdot \bfsn &= 0 \qquad {\rm on} \quad  z = -h \\
    \nabla \phi \cdot \bfsn &= 1 \qquad {\rm on} \quad  x = \pm a 
\end{aligned}
\right.
\label{eqn:extBc}
\end{equation}
where $\phi$ is the velocity potential field, $\omega$ is frequency, $g$ is the acceleration due to gravity and $\mathbf{n}$ is the unit outward normal vector. In addition to the above boundary condition, on the porous baffle, the following Darcy's law is enforced
\begin{equation}
    \nabla \phi \cdot \bfsn = - \nabla \phi \cdot \bfsn = i \sigma [ \phi^+ - \phi^-] 
\label{eq:DarcyLinear}
\end{equation}
where $\phi^+$ and $\phi^-$ are the velocity potential on either sides of the thin porous baffle. \Cref{eq:DarcyLinear} indicates that the mass flux across the porous baffle and the flow velocity across the porous baffle is linearly proportional to the pressure difference between the baffle sides and the constant $\sigma$ is referred to as porous-effect parameter that is related to the porosity parameter by:
\begin{equation}
   \sigma = \dfrac{k_1}{2\pi}b
   \label{eqn:poroparam}
\end{equation}
where $k_1$ is the wave number obtained by solving the following dispersion relation:
\begin{equation}
    k_1 \tanh(k_1h) = \dfrac{\omega^2}{g}
\end{equation}
In \Cref{eqn:poroparam}, $b=0$   corresponds to an impermeable baffle, whereas $b\rightarrow \infty$ refers to an infinitely permeable baffle (i.e., absence of baffle). Cho and Kim~\cite{chokim2008}, based on experiments, derived the following empirical relation between the actual porosity, $P$ and the porosity parameter, $b$:
\begin{equation}
    b = 57.63 P - 0.9717, \quad P \in [0.05,0.4]
\end{equation}

The wave elevation $(\zeta(x,t)= Re\{\eta(x)e^{-i\omega t}\}) $ on tank wall can be described as,
\begin {equation}
     \eta(x) = \frac{\omega^2 \xi}{g} \phi(x,0)
\end{equation}
Then, the normalized amplification factor is,
\begin{equation}
   \overline{\eta}= \frac{\eta}{\xi}= \frac{\omega^2 }{g} \phi(x,0)
    \label{amp}
\end{equation}
Sloshing pressure $(P(-a,z,t)= Re\{p(-a,z)e^{-i\omega t}\})$ on tank wall is,
\begin{equation}
     p(-a,z)= \rho \omega^2 \xi \int_{-h}^{0} \phi(-a,z)
\end{equation}

The sloshing force can be obtained by integrating the sloshing pressure
\begin{equation}
     f= \rho \omega^2 \xi \int_{-h}^{0} \phi(-a,z) \,dz
\end{equation}
 The normalized sloshing force is, 
\begin{equation}
    \overline{F}= \frac{|f|}{\xi \rho g}
    \label{slosh_force}
\end{equation}
The normalized frequency is,
\begin{equation}
    \overline{\omega} = \frac{\omega^2h}{g}
\end{equation}
\begin{figure}[htpb]
\subfigure[Case-A]{\includegraphics[scale=0.18]{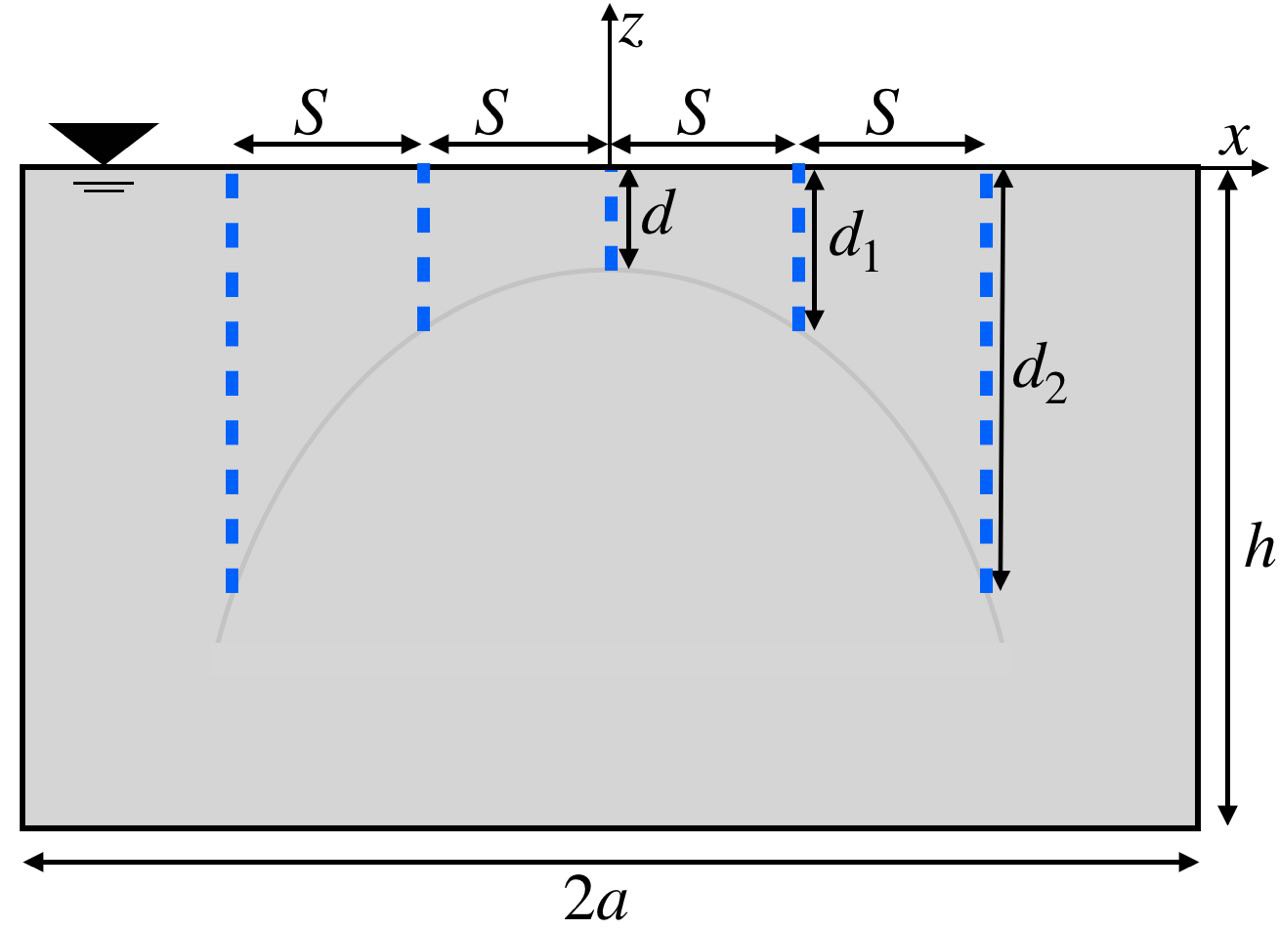}}
\subfigure[Case-B]{\includegraphics[scale=0.24]{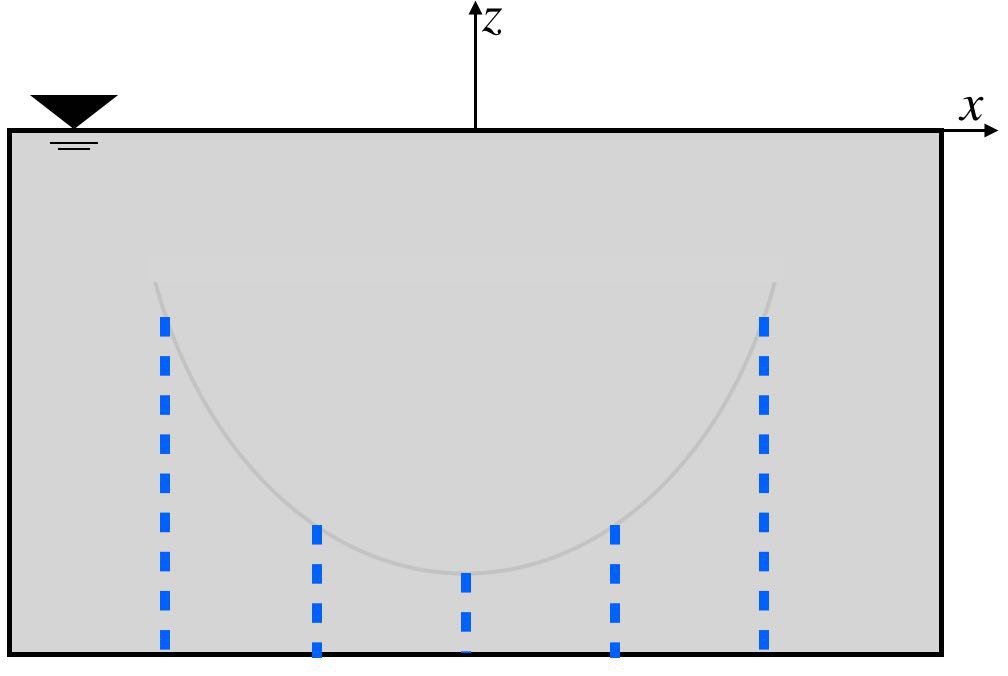}}

\subfigure[Case-C]{\includegraphics[scale=0.23]{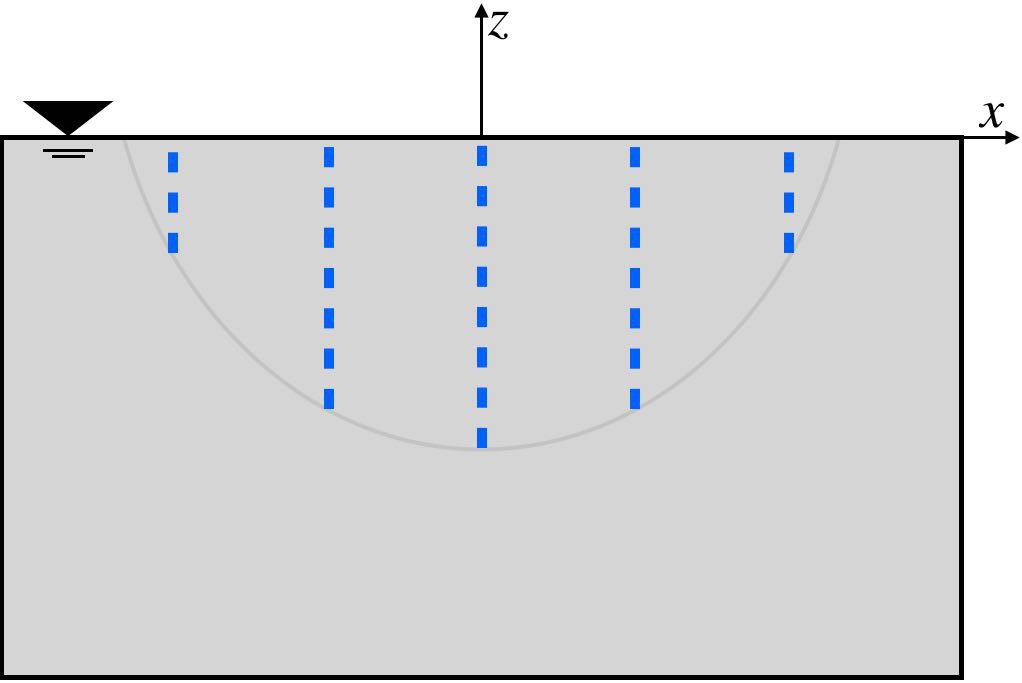}}
\subfigure[Case-D]{\includegraphics[scale=0.19]{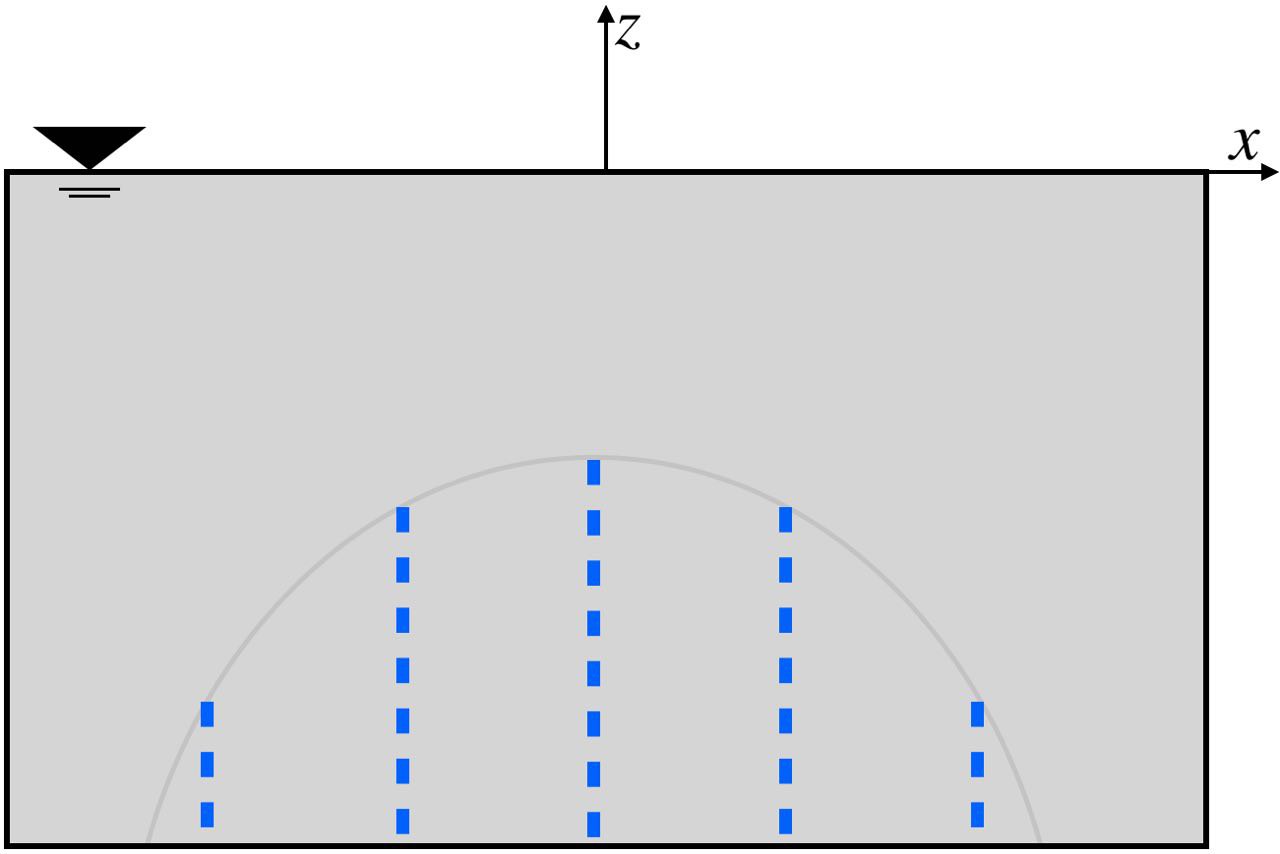}}
\caption{Schematic representation of rectangular tank with five (a,c) top mounted vertical baffles (b,d) bottom mounted vertical baffles, where the endpoints of the baffles follow a parabolic profile.}
\label{fig:schematic_top_para}
\end{figure}
\begin{table}[htpb]
    \centering
    \baselinestretch{3}
    \caption{Details of the difference configurations considered in this study, c.f.~\Cref{fig:schematic_top_para}.}
    \begin{tabular}{cllcccc}
    \hline 
         Load case  &  Baffle arrangement & Equation of parabola & $d/h$ & $d_1/h$ & $d_2/h$ & $\alpha$ value \\
         \hline
         Case-A & Top-mounted - concave & $x^2 = -4\alpha (y+d)$ & 0.4 & 0.5 & 0.8 & 3.6 \\
         Case-B  & Bottom-mounted - concave & $x^2 = 4\alpha (y+h-d)$ & 0.4 & 0.5 & 0.8 & 3.6 \\
         Case-C & Top-mounted - convex & $x^2 = 4\alpha (y+d)$ & 0.8 & 0.7 & 0.4 & 3.6 \\
         Case-D & Bottom-mounted - convex & $x^2 = -4\alpha (y+h-d)$ & 0.8 & 0.7 & 0.4 & 3.6 \\
         \hline
    \end{tabular}
    \label{tab:rec_tank_para_baffle_conve}
\end{table}

\section{Overview of the scaled boundary finite element method}
\label{sec:sbfemintro}
\begin{figure}[htpb]
\centering 
\includegraphics[scale=0.7]{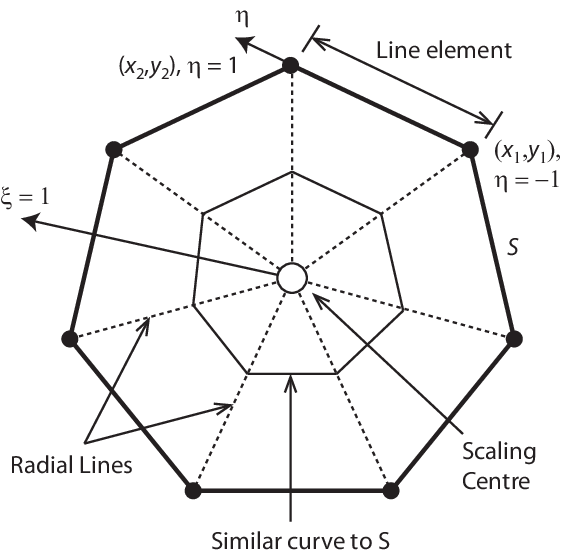}
\caption{Scaled boundary finite element method - description of scaled boundary coordinates}
\label{fig:sbfemdescrip}
\end{figure}
The SBFEM relies on defining a \emph{scaling centre}, as shown in  \Cref{fig:sbfemdescrip}, from which the entire domain is directly visible. This method reduces the governing partial differential equations to a set of ordinary differential equations; this is done by introducing a local radial-circumferential coordinate system and variational framework. Based on this, in the radial direction, smooth analytical functions are used to define the solution while numerical solutions are sought around the boundary. Within the framework of the scaled boundary finite element method, the geometry of the boundary is described by interpolating the points on the boundary using shape functions as:
\begin{equation}
\xx(\xi,\eta) = (x(\xi,\eta), y(\xi,\eta)) = \xi \bfn(\eta) \xx_I
\label{eqn:sbfemcoord}
\end{equation}
where $\xi$ is the radial coordinate with $\xi=$ 1 on the boundary and $\xi=$ 0 at the scaling center (see \Cref{fig:sbfemdescrip}), $\xx_I$ are the coordinates on the boundary and $\bfn(\eta)$ is the shape function matrix. Following the general concept of the SBFEM~\cite{songwolf2000,songwolf2000a}, the gradient operator in \Cref{eqn:governeqn} is written as:
\begin{equation}
\nabla = \left[ J(\eta) \right]^{-1} \left[ \begin{array}{cc} 1 & 0 \\ 0 & \xi \end{array} \right] \left\{ \begin{array}{c} \frac{\partial}{\partial \xi} \\ \frac{\partial}{\partial \eta} \end{array} \right\} = \mathbf{b}^1(\eta) \frac{\partial}{\partial \xi} + \frac{1}{\xi} \mathbf{b}^2(\eta) \frac{\partial}{\partial \eta},
\label{eqn:sbfemgradop}
\end{equation}
where the Jacobian is given by:
\begin{equation}
\left[ J(\eta) \right] = \left[ \begin{array}{cc} x & y \\ x_{,\eta} & y_{,\eta} \end{array} \right]
\end{equation}
where the subscript refers partial derivative with respect to the circumferential coordinate, $\eta$. The vectors $\mathbf{b}^1(\eta)$ and $\mathbf{b}^2(\eta)$ in \Cref{eqn:sbfemgradop} can be expressed as:
\begin{equation}
\mathbf{b}^1(\eta) = \frac{1}{|J|} \left\{ \begin{array}{c} y_{,\eta} \\ -x_{,\eta} \end{array} \right\}, \hspace{5pt} \mathbf{b}^2(\eta) =  \frac{1}{|J|} \left\{ \begin{array}{c} -y\\ x \end{array} \right\}
\end{equation}
where $|J|$ is the determinant of the Jacobian. The potential at a point in a polygon is approximated using the same shape functions as used for the geometry:
\begin{equation}
\phi(\xi,\eta) = \bfn(\eta) \phi(\xi) 
\label{eqn:sbfempress}
\end{equation}
and by employing the variational method, the following partial differential equation for the nodal pressure amplitude is obtained:
\begin{equation}
\mathbf{E}_0 \xi^2 \phi(\xi)_{,\xi \xi} + ( \mathbf{E}_0 - \mathbf{E}_1 + \mathbf{E}_1^{\rm T} ) \xi \phi(\xi)_{,\xi} - \mathbf{E}_2 \phi(\xi) = \mathbf{0}
\label{eqn:sbfemgoverneqn}
\end{equation}
where $\mathbf{E}_0, \mathbf{E}_1$ and $\mathbf{E}_2$ denote the scaled boundary coefficient matrices calculated element-by-element and assembled using standard finite element techniques. The coefficient matrices are given by:
\begin{equation}
\mathbf{E}_0 =\int\limits_\eta \mathbf{B}_1^T \mathbf{B}_1~|J|~\mathrm{d}\eta, \quad 
\mathbf{E}_1 =\int\limits_\eta \mathbf{B}_2^T \mathbf{B}_1~|J|~\mathrm{d}\eta, \quad
\mathbf{E}_2 =\int\limits_\eta \mathbf{B}_2^T \mathbf{B}_2~|J|~\mathrm{d}\eta, 
\end{equation}
where,
\begin{equation*}
\mathbf{B}_1 = \mathbf{b}^1 \bfn{(\eta)}, ~~~ \mathbf{B}_2 = \mathbf{b}^2 \bfn(\eta)_{,\eta}
\end{equation*}
Note that the scaled boundary coefficient matrices are evaluated over the element (or subdomain) boundaries. The corresponding stiffness matrix can be obtained by converting \Cref{eqn:sbfemgoverneqn} into a first order differential equation having twice the number of unknowns, this is done by introducing, $\mathbf{q}^h(\xi)$ such that:
\begin{equation}
    \xi \left\{ \begin{array}{c} \mathbf{\phi}^h(\xi) \\ \mathbf{q}^h(\xi) \end{array} \right\}_{,\xi} = - \mathbf{Z} \left\{ \begin{array}{c} \mathbf{\phi}^h(\xi) \\ \mathbf{q}^h(\xi) \end{array} \right\},
\end{equation}
where $\mathbf{Z}$ is the Hamiltonian matrix given by:
\begin{equation} \label{EQ37}
\mathbf{Z} =\left[\begin{array}{cc}
-\mathbf{E}_0^{-1} \mathbf{E}_1^{\rm T} & \mathbf{E}_0^{-1} \\
\mathbf{E}_2 - \mathbf{E}_1 \mathbf{E}_0^{-1}\mathbf{E}_{1}^{\rm T} & \mathbf{E}_1 \mathbf{E}_{0}^{-1}
\end{array}\right],
\end{equation}
and the nodal flux vector is given by:
\begin{equation}
    \mathbf{q}^h(\xi) = \mathbf{E}_0 \xi \mathbf{\phi}^h(\xi)_{,\xi} + \mathbf{E}_1^T \mathbf{\phi}^h(\xi).
\end{equation}
The unknown potential field $\mathbf{\phi}^h(\xi)$ is obtained by performing an eigenvalue decomposition of the Hamiltonian matrix, resulting in:
\begin{equation}
    \mathbf{Z} \bpsi = \bpsi \bl ,
\end{equation}
with $\bpsi$ containing the eigenvectors and $\bl$ is a matrix that contains the eigenvalues of the Hamiltonian matrix, that are partitioned as:
\begin{subequations}
\begin{align}
    \bpsi &= \left[ \begin{array}{cc} \bpsi_n^{(t)} & \bpsi_p^{(t)} \\ \bpsi_n^{(q)} & \bpsi_p^{(q)} \end{array} \right], \\
    \bl &= \left[ \begin{array}{cc} \bl_n & \mathbf{0} \\ \mathbf{0} & \bl_p \end{array} \right],
\end{align}
\end{subequations}
where the superscripts $(t)$ and $(q)$ represents the eigenvectors corresponding to the unknown potential and the gradient of the potential, respectively and the subscripts $n,p$ represents the negative and positive eigenvalues of the $\mathbf{Z}$ matrix. For bounded polygons, such as those considered in this paper, only the first $n$ negative eigenvalues in $\bl$ with corresponding eigenvectors in $\bpsi$ is considered. This ensures that the solution obtained at the scaling center is finite and bounded. Denoting the negative eigenvalues and eigenvectors as $\bl_n$ and $\bpsi_n$, respectively, the potential field is given by:
\begin{subequations}
\begin{align}
    \mathbf{\phi}^h(\xi) &= \bpsi_n^{(t)} \xi^{-\bl_n} \mathbf{c} \label{eqn:tempsol},\\
    \mathbf{q}^h(\xi) &= \bpsi_n^{(q)} \xi^{-\bl_n} \mathbf{c} \label{eqn:fluxsol},
\end{align}
\end{subequations}
where $\mathbf{c}$ in \cref{eqn:tempsol} are the integration constants that are determined from the nodal solution of the potential field, i.e., $\mathbf{\phi}_b = \mathbf{\phi}^h(\xi=1)$ as:
\begin{equation}
    \mathbf{c} = \left( \bpsi_n^{(t)} \right)^{-1} \mathbf{\phi}_b
    \label{eqn:consol}.
\end{equation}
The nodal force vector $\mathbf{F} = \mathbf{q}(\xi=1)$, can be computed by setting $\xi=$1 and substituting \cref{eqn:consol} into \cref{eqn:fluxsol} as:
\begin{equation}
    \mathbf{F}^e = \bpsi_n^{(q)} \left( \bpsi_n^{(t)}\right)^{-1} \mathbf{\phi}_b
    \label{eqn:nodalforce},
\end{equation}
and the potential field in terms of the nodal unknowns $\mathbf{\phi}_b$ is given by:
\begin{equation}
    \mathbf{\phi}^h(\xi) = \bpsi_n^{(t)} \xi^{-\bl_n} \left( \bpsi_n^{(t)}\right)^{-1} \mathbf{\phi}_b
    \label{eqn:sbfemtemp}.
\end{equation}
The stiffness matrix of the polygon is computed from the eigenvectors of the Hamiltonian matrix as:
\begin{equation}
    \mathbf{K}^e = \bpsi_n^{(q)} \left( \bpsi_n^{(t)} \right)^{-1}
    \label{eqn:sbfemstiff}.
\end{equation}

\section{Numerical examples}
\label{sec:numexam}
In this section, the accuracy and robustness of the proposed framework is numerically studied. The computational domain, i.e., rectangular tank with baffles is `minimally' divided into subdomains such that the star convexity is met as required by the SBFEM. On the boundary of the subdomain, the polynomial order of the shape functions is varied and its influence on the accuracy of the solution is studied. Subsequently, the focus of the present investigations on the effect of configuration, porosity, tank width, submergence depth of the baffles, and the spacing between the baffles are parametrically explored. Various results are presented, such as the amplification factor and sloshing loads on the tank walls. An in-house code is developed in Matlab, and all simulations were performed on a 64 GB RAM, Intel Core i9-10900X $\times$ 20 processors with 3.7 GHz. 

\subsection{Convergence study }
\begin{figure}[htpb]
\centering
\includegraphics[scale=0.3]{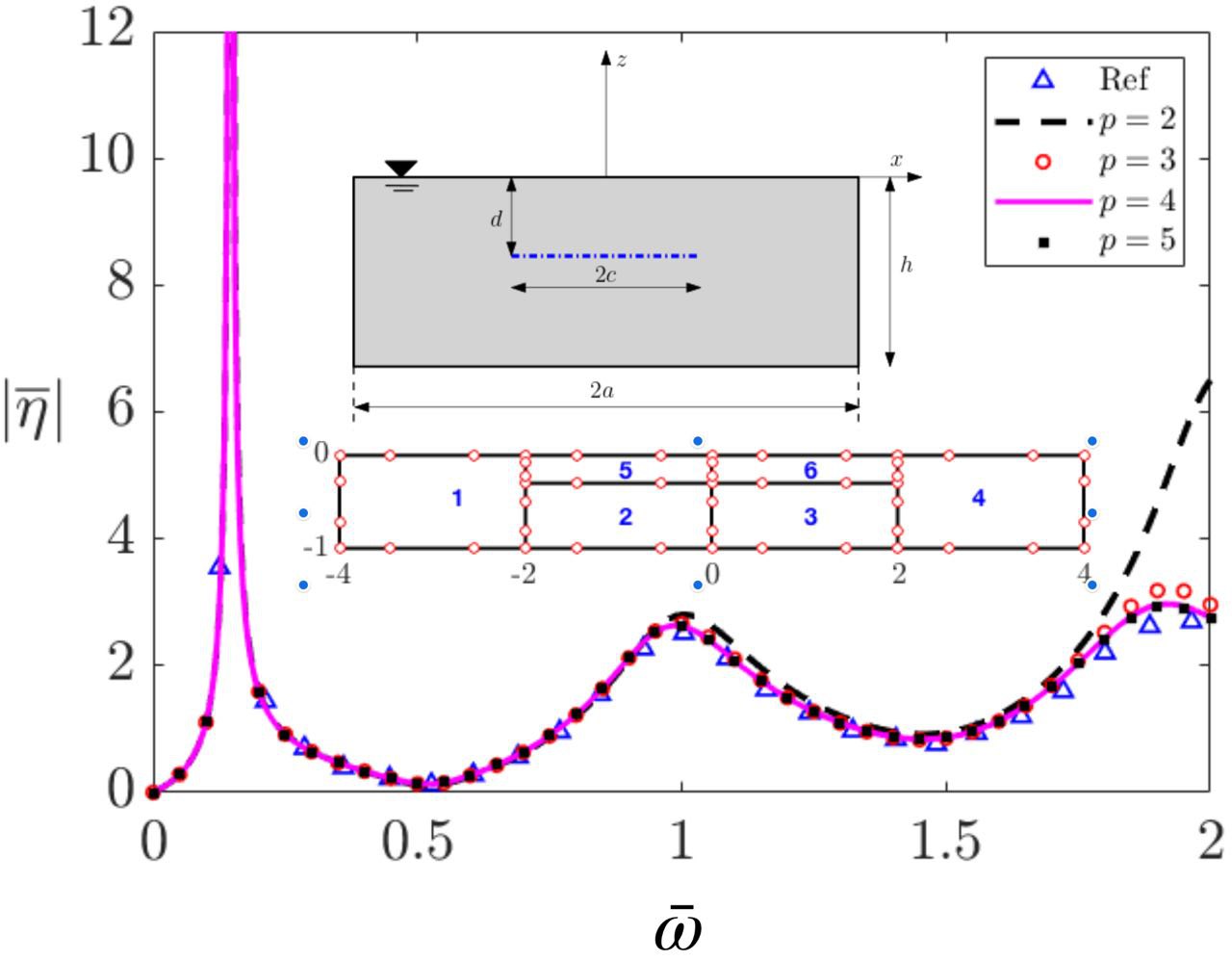}
\caption{Rectangular tank with horizontal porous baffle: convergence of amplification factor with order of element, $p$ used to discretize.}
\label{fig:rectang_mesh_conv}
\end{figure}

Before proceeding with a numerical study on the influence of various parameters on the sloshing dynamics, the accuracy and convergence properties of the proposed framework is studied for a swaying rectangular tank with a submerged horizontal baffle \cite{chochoi2017}. The physical parameters of the test setup are $a/h=4, d/h=0.1, c/h= 2.0, P=0.3$. For brevity, the configuration is shown in the inset of \Cref{fig:rectang_mesh_conv}. The presence of the baffle introduces discontinuity in the potential field, and the domain does not satisfy star convexity. To satisfy star convexity, the domain is divided into six scaled boundary subdomains. Each edge is discretized with 3$^{\rm rd}$ order Lagrange elements.  `\textcolor{red}{Red}' circles represent the nodes, and the `\textcolor{blue}{blue}' text refers to the sub-domain numbers in the schematic representation of the nodes on the boundary of each subdomain as shown inset of \Cref{fig:rectang_mesh_conv}. The order of the approximation functions on the boundary of the domain is increased and the first three fundamental frequencies are computed for porosity value $0.3$. The results from the present work are compared against the results available from Cho et al.~\cite{chochoi2017}, and the same is tabulated in \Cref{tab:rec_tank_no_baffle_conve}. The normalized amplification factor $(\overline{\eta})$ as a  function of non-dimensional frequency $(\overline{\omega})$ for different polynomial order is shown in \Cref{fig:rectang_mesh_conv}. It is seen that as the order of the polynomial is increased, the accuracy is improved and that a fifth order polynomial is enough to yield accurate results, this results in only 94 degrees of freedom.

\begin{table}[htpb]
    \centering
    \caption{Convergence of the first three fundamental frequency of a rectangular tank with a center horizontal porous baffle with porosity  $P=0.3$.}
    \begin{tabular}{lcrrr}
    \hline
         order of element ($p$) & number of nodes & $\hat{\omega}_1$ & $\hat{\omega}_2$ & $\hat{\omega}_3$ \\
         \hline
         2 & 34 & 0.1471 & 0.9870 & --\\
         3 & 54 & 0.1471 & 0.9753 & 1.9031 \\
         4 & 74 & 0.1471 & 0.9742 & 1.8888 \\
         5 & 94 & 0.1471 & 0.9742 & 1.8888\\
         Ref.~\cite{chochoi2017} & -- & 0.1467 & 0.9741 & 1.8876 \\
         \hline
    \end{tabular}
    \label{tab:rec_tank_no_baffle_conve}
\end{table}

\subsection{Validation}
\begin{figure}[htpb]
    \subfigure[]{\includegraphics[scale=0.19]{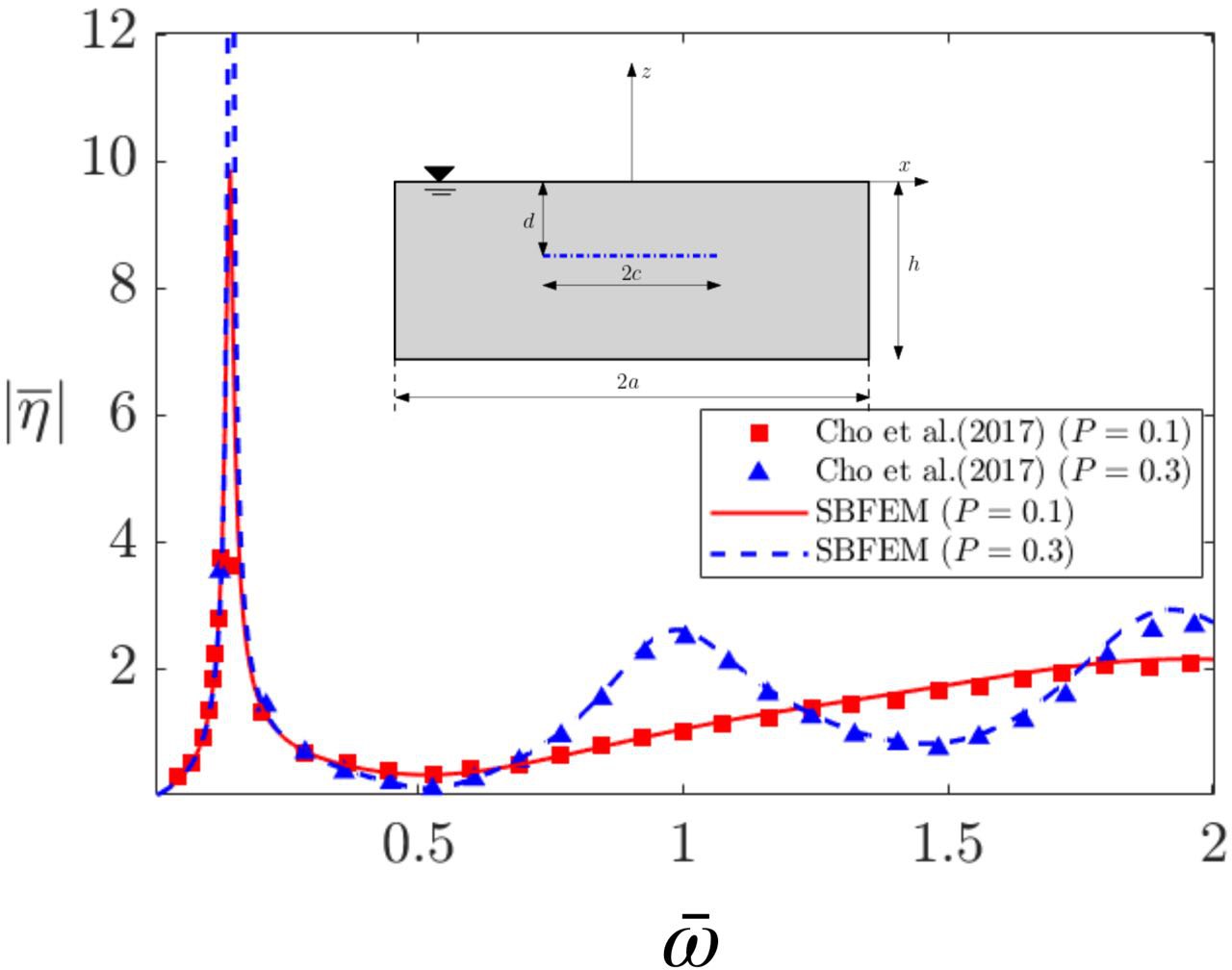}}
   \subfigure[]{\includegraphics[scale=0.185]{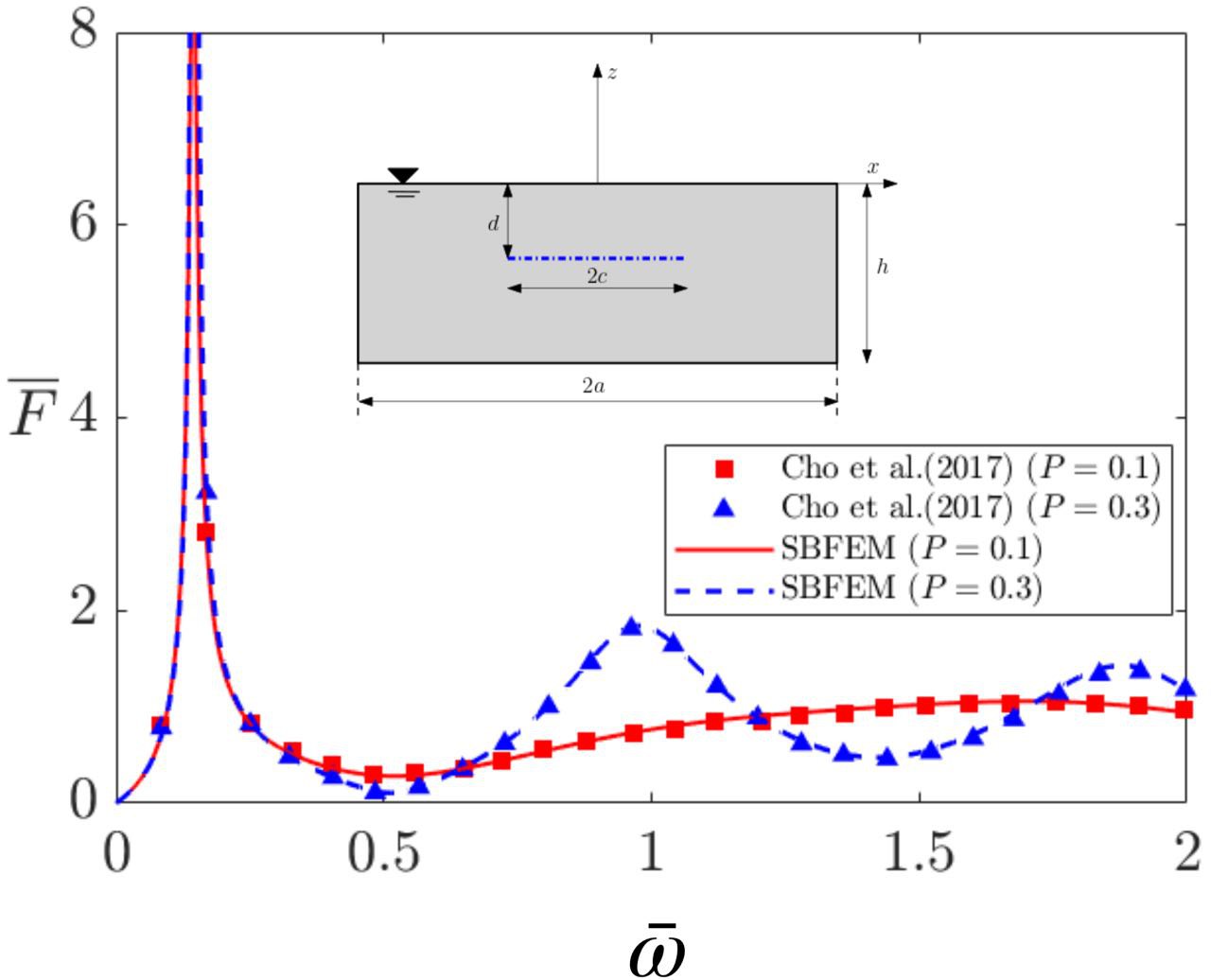}}
    \caption{(a) Amplification factor, $|\overline{\eta}|$ and (b) Normalized sloshing force, $\overline{F}$as a function of normalized frequency $\overline{\omega}$ for different porosity. The results from the present framework is compared with the results Cho et al. based on boundary element method~\cite{chochoi2017}.}
    \label{fig:rectan_valid}
\end{figure}

To ensure the integrity of the developed numerical model, the computed numerical results are compared with results available in the literature. \Cref{fig:rectan_valid} shows the normalized amplification factor $(\overline{\eta})$ and sloshing force $(\overline{F})$ as a function of normalized frequency, $(\overline{\omega})$. A very good agreement with the results obtained from the boundary element method~\cite{chochoi2017} is seen. For both $x=-a$ and $x=a$, the results are identical, consistent with the findings of Cho et al. It is noteworthy to mention that the matrix size of the model is a mere $94$, which is nearly $1/3$ of the matrix size of the Cho et al. ~\cite{chochoi2017}. In the upcoming sections, the results of the proposed study are presented and discussed.

\subsection{Effect of configuaration}
\begin{figure}[htpb]
\subfigure[]{\includegraphics[scale=0.5]{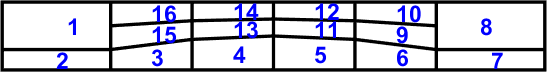}}\qquad \qquad
\subfigure[]{\includegraphics[scale=0.5]{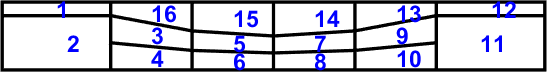}}

\subfigure[]{\includegraphics[scale=0.5]{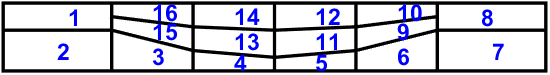}}\qquad \qquad
\subfigure[]{\includegraphics[scale=0.5]{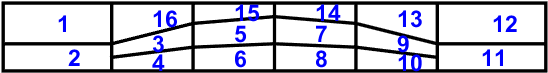}}
\caption{Rectangular tank with five (a,c) top mounted vertical baffles (b,d) bottom mounted vertical baffles, where the endpoints of the baffles form a parabolic curve divided into sixteen scaled boundary subdomains.}
\label{fig:meshdetails_top_para}
\end{figure}

\begin{figure}[htpb]
\subfigure[]{\includegraphics[scale=0.175]{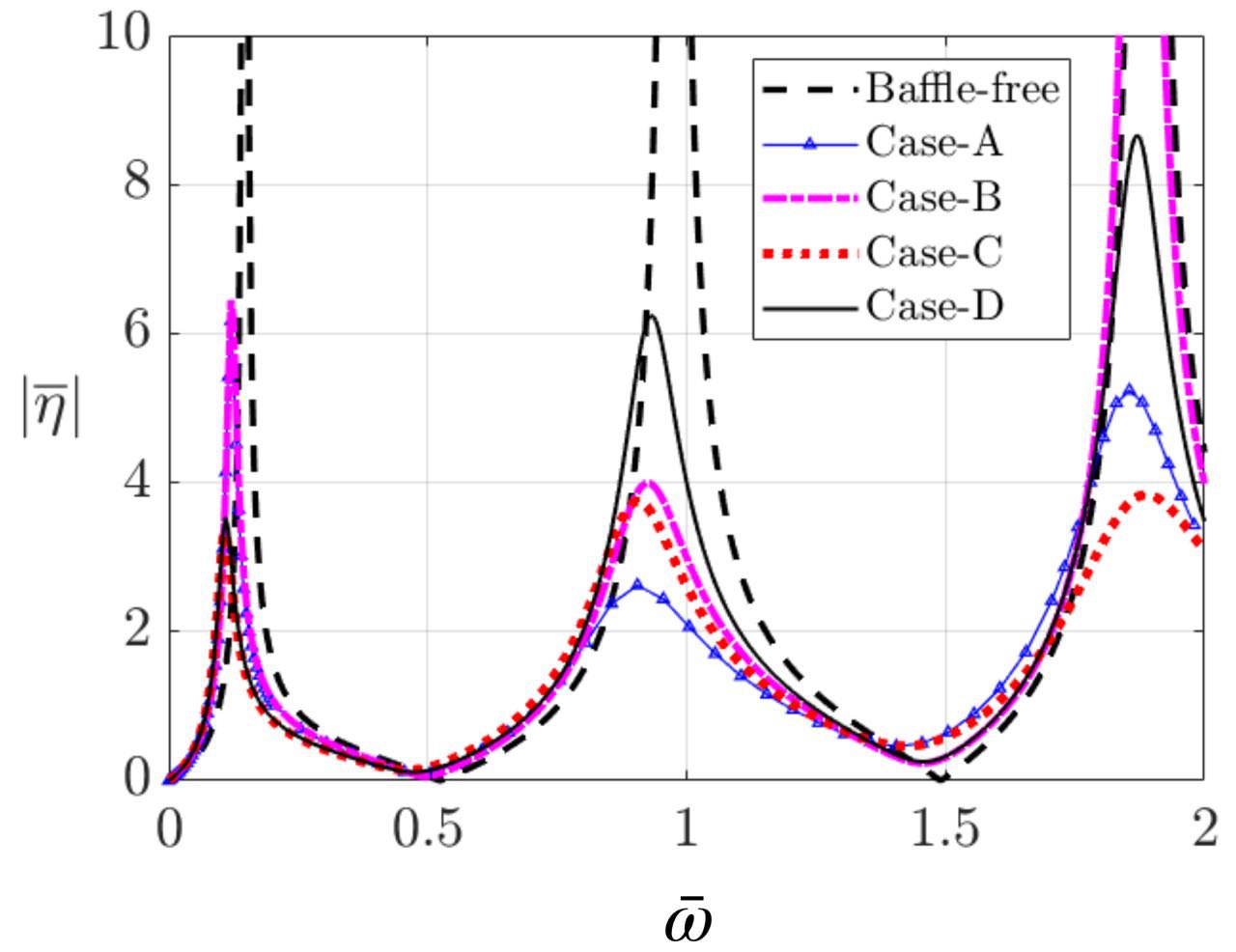}}
\subfigure[]{\includegraphics[scale=0.175]{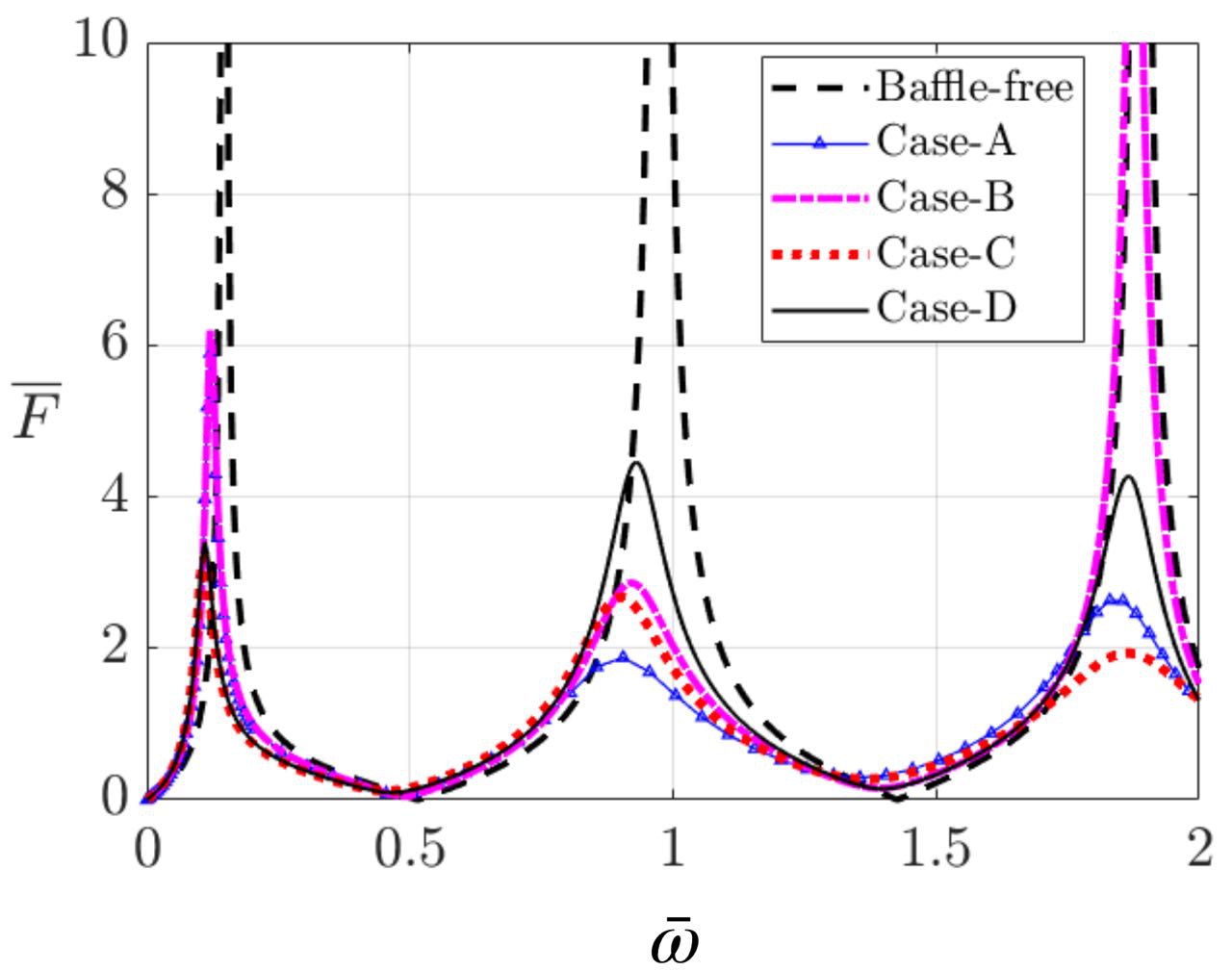}}
\caption{(a,b) Comparison of amplification factor at $x=-a$ and normalized sloshing force on tank wall for different baffle configuration.}
\label{fig:rect_tank_toppara_effect}
\end{figure}

This section is devoted to investigating the best configuration among the four, as shown in \Cref{fig:schematic_top_para}. The physical parameters are given in \Cref{tab:rec_tank_para_baffle_conve} in which the spacing between the baffle is maintained the same in all the cases $S/a = 0.3$. Each tank is divided into $16$ scaled boundary subdomains to satisfy star convexity based on their geometry described in \Cref{fig:meshdetails_top_para} (a-d).
From \Cref{fig:rect_tank_toppara_effect}, it is observed that the first, third, and fifth modes of the natural frequency exhibit resonance peaks in both non-baffle (i.e., baffle-free tank) and baffle configurations. In the presence of the baffle, the peak amplitude is effectively reduced for the first and third modes across all cases. A similar trend is observed for the fifth mode, except in case-B. It is noticeable that among the four cases, the resonance peak amplitude for the first and third modes of the non-dimensional frequency is lower in cases A and C compared to cases B and D. Moreover, for the fifth mode, this difference becomes more pronounced, with the peak amplitude in cases A and C being more than $50\%$ lower than in cases B and D. This effectiveness is attributed to the enhanced energy dissipation due to the direct interaction of the top-mounted baffles with the free surface motion. In contrast, bottom-mounted baffles primarily affect the flow near the base, leading to weaker attenuation of surface waves and less effective suppression of sloshing. For the first and fifth modes of the non-dimensional frequency, case-C exhibits a lower peak amplitude than case-A, while for the third mode, case-A performs better than case-C. Evaluating the overall performance, case-C stands out as the most effective configuration to reduce the sloshing phenomena, striking a better balance across all modes. In the subsequent sections, the influence of baffle porosity $(P)$, tank width $(a)$, submergence depth $(d,d_2)$, and the spacing between two porous baffles $(S)$ on the amplification factor and normalized sloshing force at $x=-a$ is examined for case-C.

\subsection{Effect of porosity}

\begin{figure}[htpb]
\subfigure[]{\includegraphics[scale=0.175]{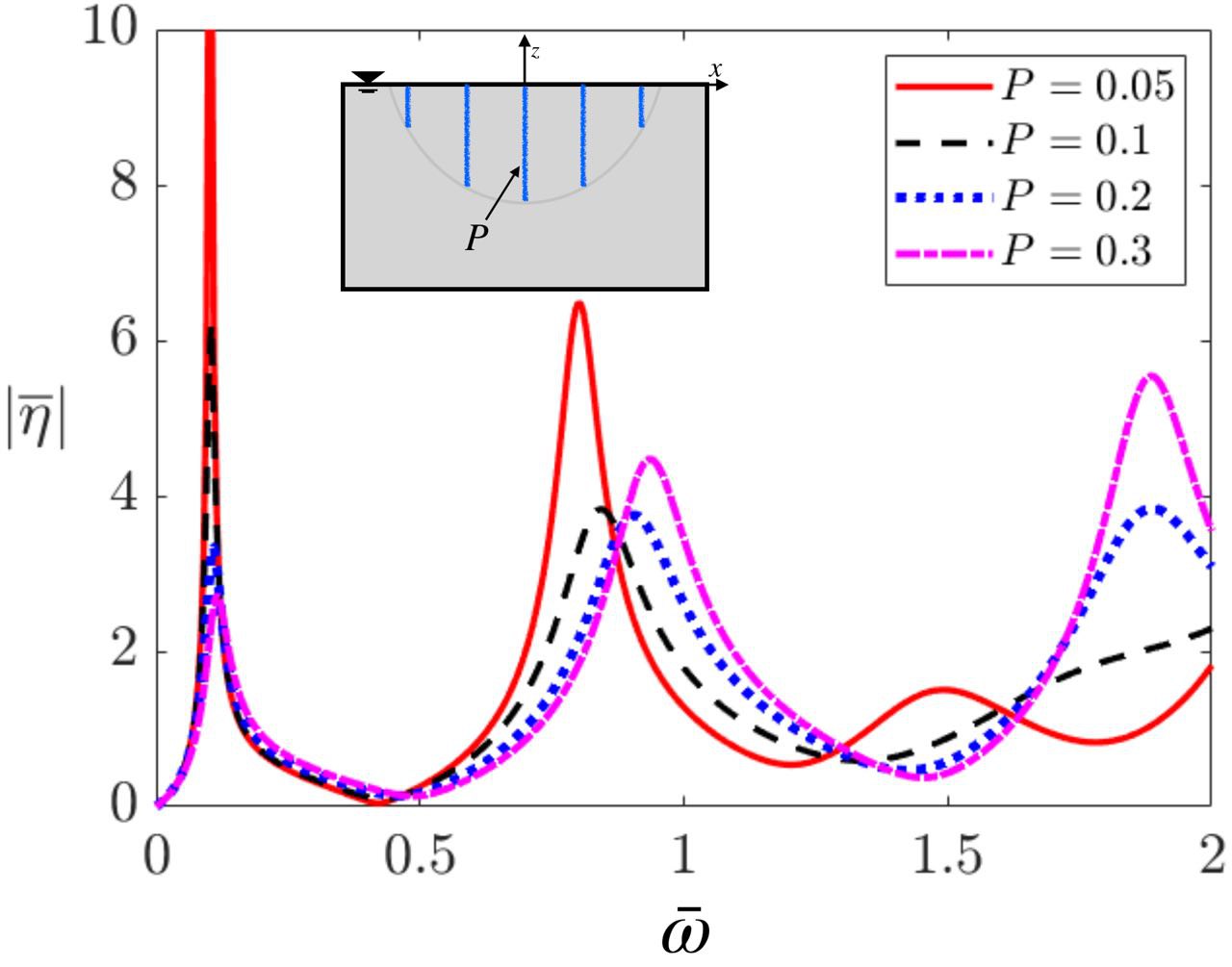}}
\subfigure[]{\includegraphics[scale=0.18]{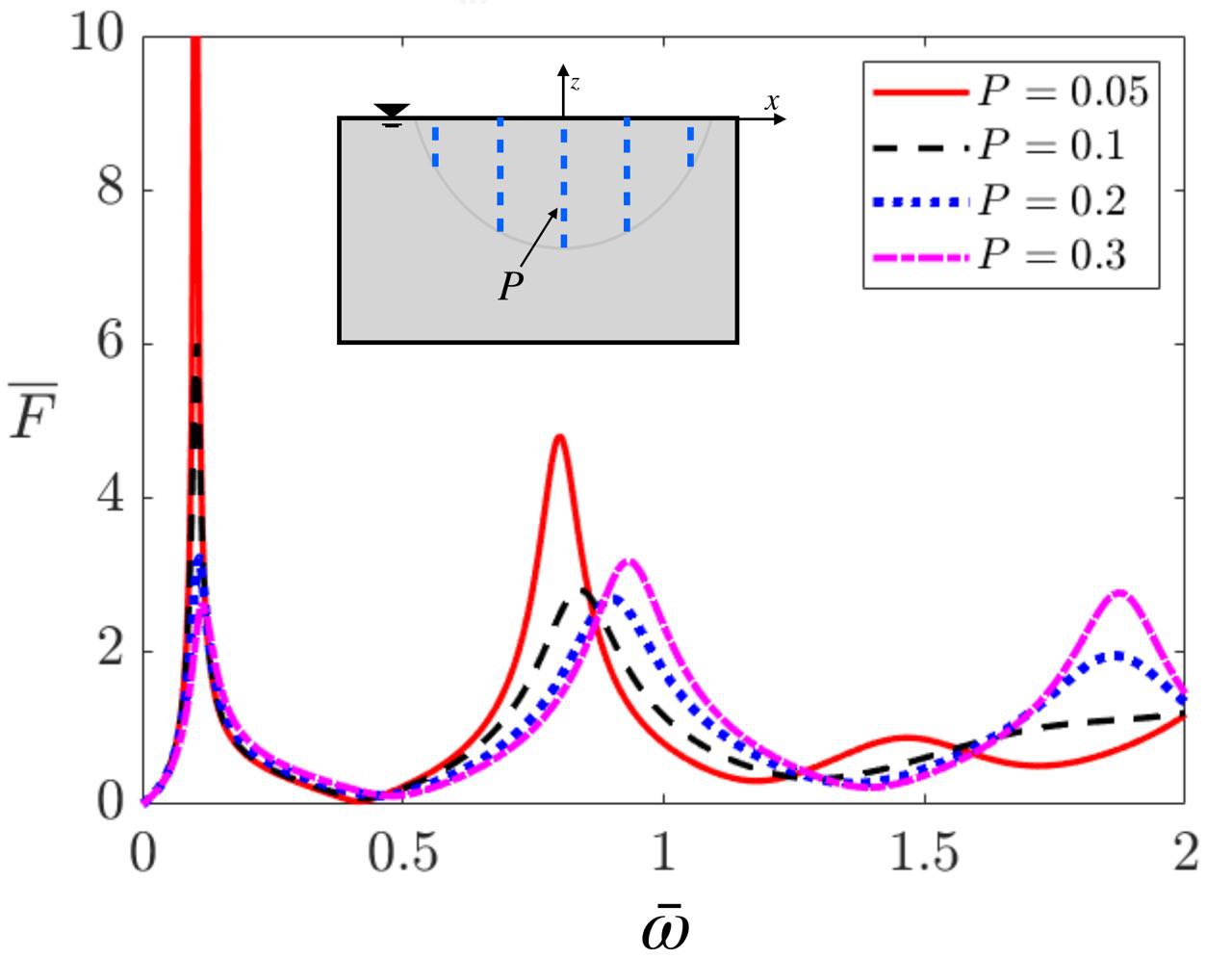}}
\caption{(a,b) Amplification factor at $x=-a$ and normalized sloshing force on tank wall for different porosity of baffle for case (C) $(a/h = 4.0, S/a = 0.3, d/h = 0.8, d_2/h = 0.4)$. }
\label{fig:rect_tank_toppara_diff_p}
\end{figure}

\Cref{fig:rect_tank_toppara_diff_p} illustrates the variation of amplification factor and normalized sloshing force at $x=-a$ with changing the porosity of the baffle $(P= 0.05,0.1,0.2,0.3)$ for fixed values of $a/h=4.0,S/a=0.3,d/h=0.8,d_2/h=0.4$. The peak amplitudes are predominantly observed in the first, third, and fifth modes of the non-dimensional frequency, while the second and fourth modes exhibit nodes at the tank wall. The largest peak amplitude is observed at the lowest frequency, while its bandwidth is the narrowest. A substantial change in the peak of the $3$rd and $5$th modes is noticeable for porosity $0.1$ and below, whereas for porosity $0.2$ and above, the peak patterns remain preserved. The amplification factor at the fifth mode diminished notably as the porosity parameter decreased.  Overall, the results indicate that porosity values $0.1$ and $0.2$ provide the best performance in reducing sloshing amplitude since it enhances wave energy dissipation by effectively redistributing flow interactions within the tank. The pressure difference across the porous baffle primarily drives the similar trend observed in the reduction of sloshing-induced force (\Cref{fig:rect_tank_toppara_diff_p}(b)) on the tank wall for varying porosities.

\subsection{Effect of tank width}
\begin{figure}[htpb]
\subfigure[]{\includegraphics[scale=0.176]{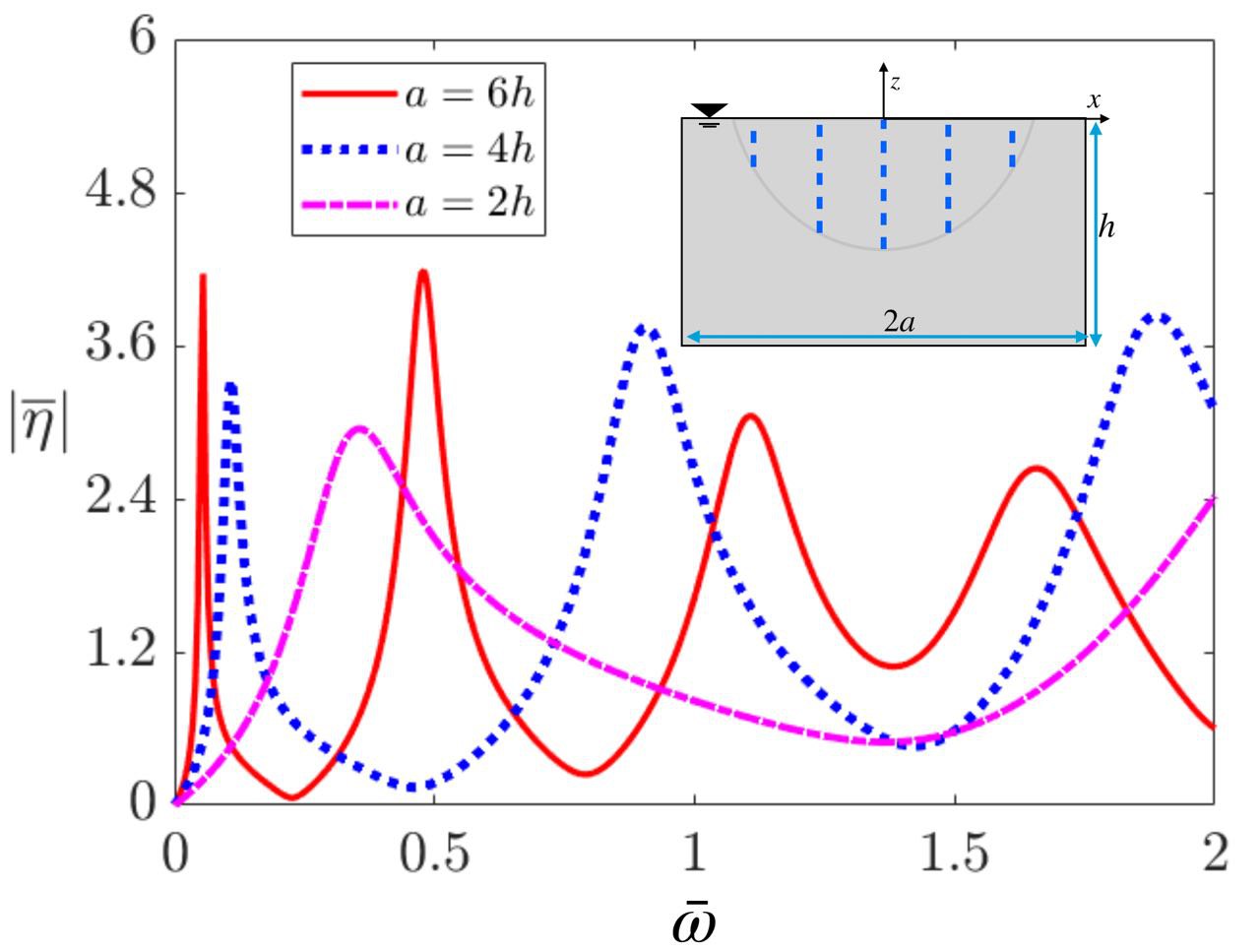}}
\subfigure[]{\includegraphics[scale=0.175]{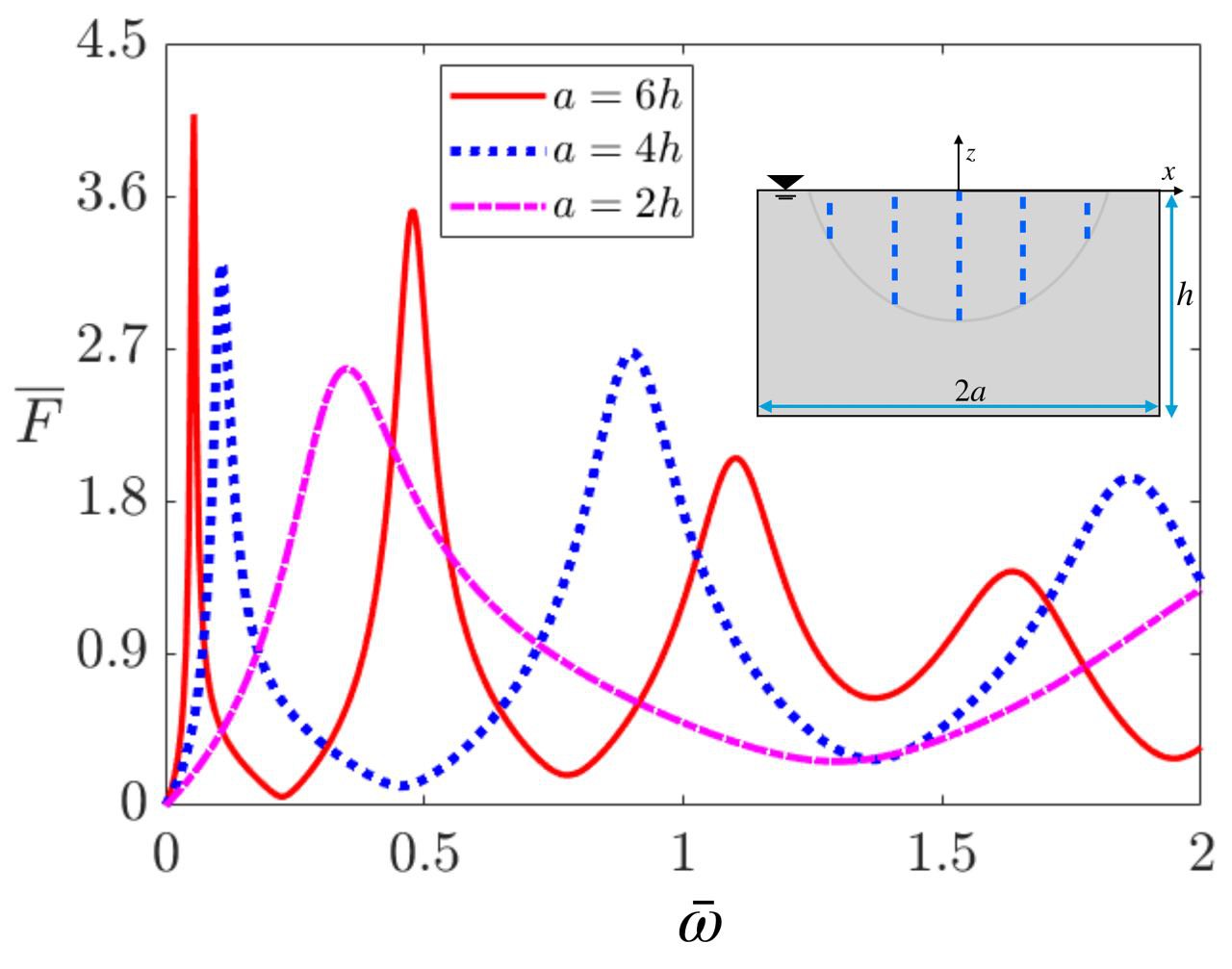}}
\caption{(a,b) Amplification factor at $x=-a$ and normalized sloshing force on tank wall for different tank width $(S/a = 0.3, P=0.2, d/h = 0.8,d_2/h = 0.4)$.}
\label{fig:rect_tank_toppara_diff_a_by_h}
\end{figure}

\Cref{fig:rect_tank_toppara_diff_a_by_h} shows the trend of the free-surface amplification factor and sloshing force on the wall when tank width is varied $(a=6h, 4h, 2h)$, whereas the other physical parameter are kept fixed $(S/a=0.3, d/h= 0.8, d_2/h=0.4, P=0.2)$. As the tank width increases, the number of sloshing modes increases, leading to a greater number of resonance peaks. Since wavelength is inversely proportional to frequency, it is physically intuitive that lower resonance frequencies will emerge. This interpretation aligns with the plots in \Cref{fig:rect_tank_toppara_diff_a_by_h}. From the figure, it is evident that among the three different values of $a/h$, the first value ($a/h=6$) exhibits the widest separation between two consecutive resonance peaks. As a result, a wider tank leads to more pronounced fluctuations due to the increased free-surface displacement. This larger displacement amplifies the system's response, resulting in a higher amplification factor. Additionally, a wider tank allows for greater free-surface wave propagation, which enhances the amplification factor and intensifies sloshing loads, exerting greater dynamic forces on the tank walls.

\subsection{Effect of submergence depth}

\begin{figure}[htpb]
\subfigure[]{\includegraphics[scale=0.176]{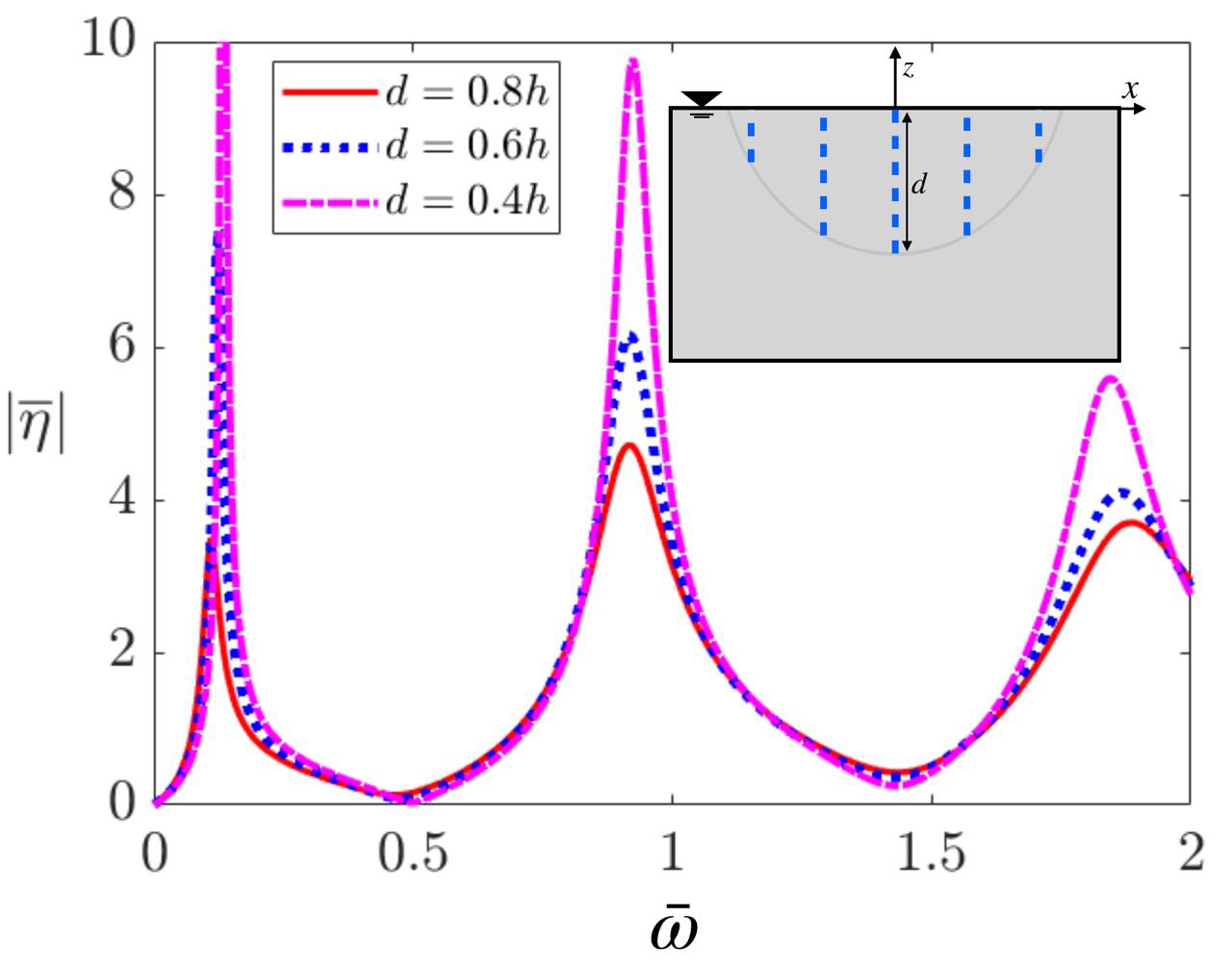}}
\subfigure[]{\includegraphics[scale=0.175]{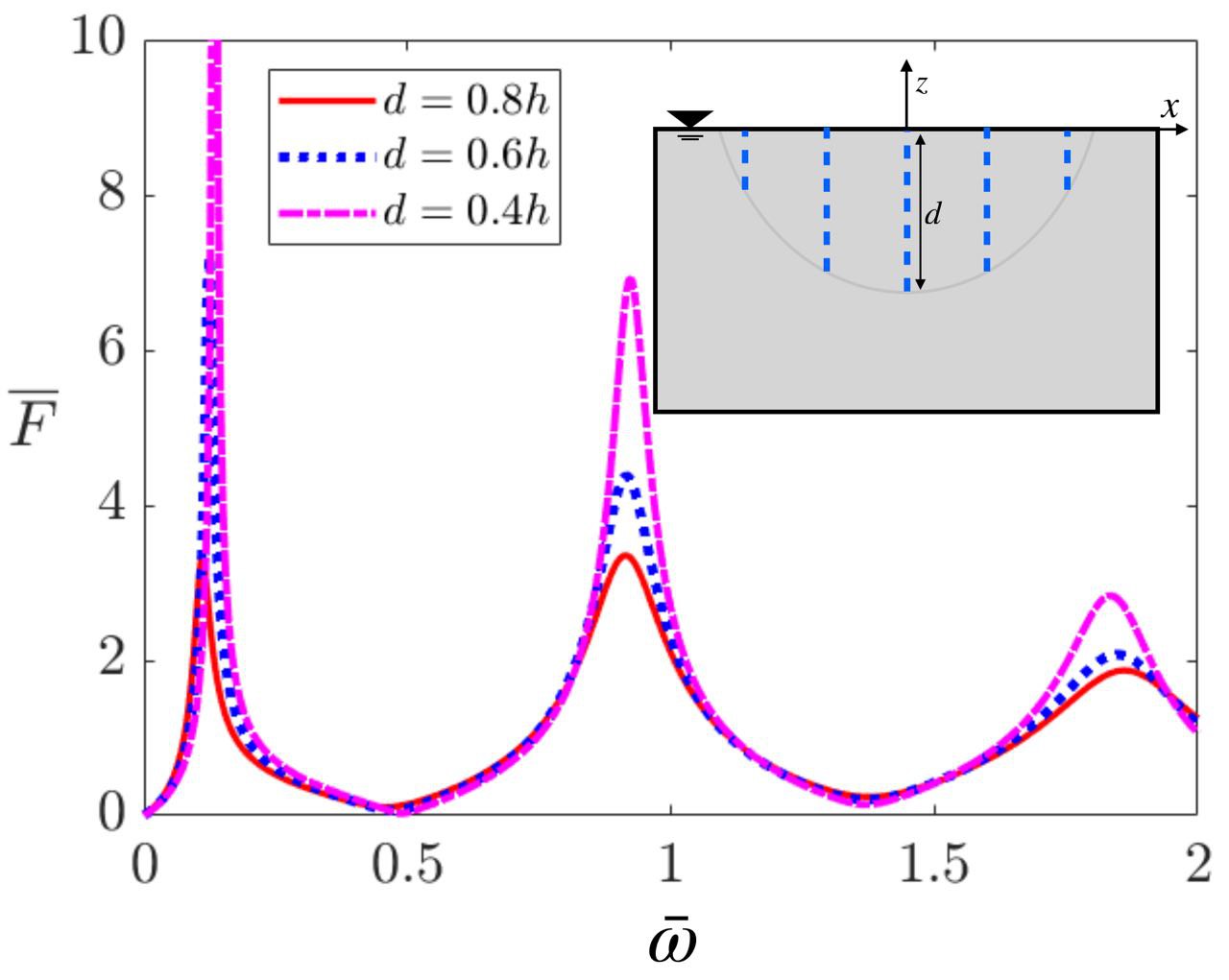}}
\caption{(a,b) Amplification factor at $x=-a$ and normalized sloshing force on tank wall for different middle baffle submergence depth $(d)$ $(a/h = 4.0, S/a = 0.3, P= 0.2, d_2/h = 0.3)$.}
\label{fig:rect_tank_toppara_diff_d}
\end{figure}

\begin{figure}[htpb]
\subfigure[]{\includegraphics[scale=0.177]{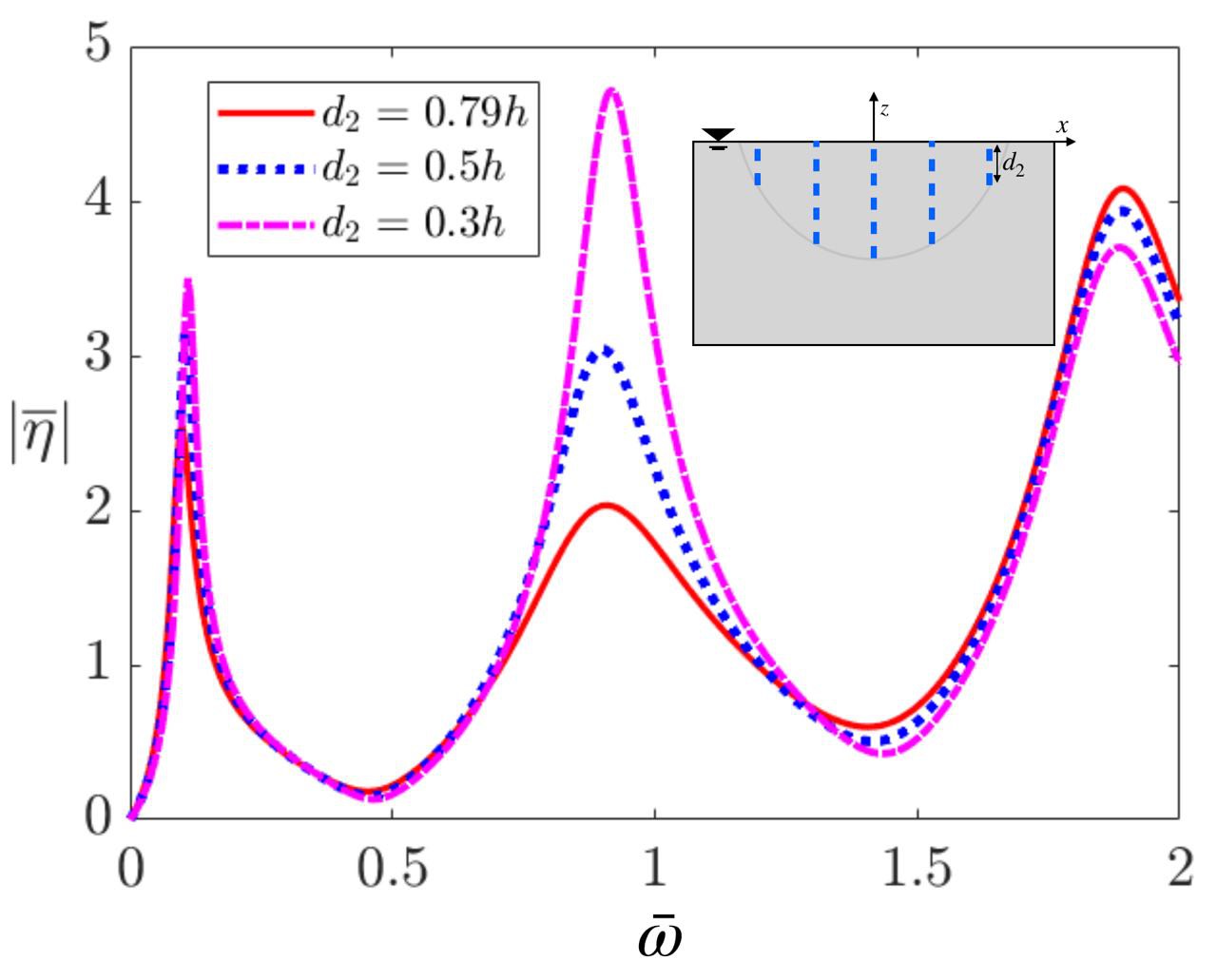}}
\subfigure[]{\includegraphics[scale=0.175]{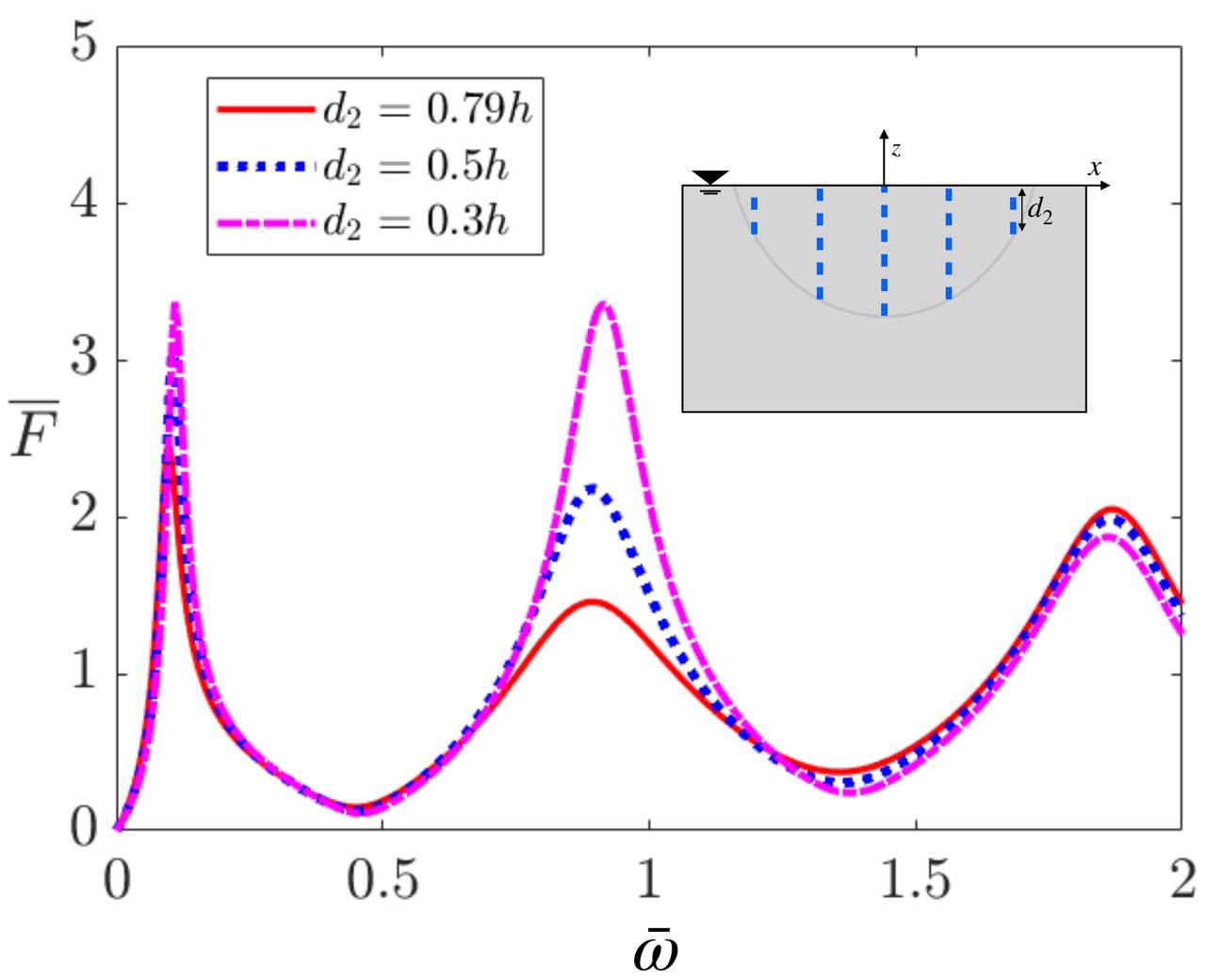}}
\caption{(a,b) Amplification factor at $x=-a$ and normalized sloshing force on tank wall for different extreme side baffle submergence depth $(d_2)$ $(a/h = 4.0, S/a = 0.3, P = 0.2, d/h = 0.8)$.}
\label{fig:rect_tank_toppara_diff_d2}
\end{figure}

The height of the baffles plays a crucial role in determining the sloshing characteristics. \Cref{fig:rect_tank_toppara_diff_d} shows the variation in amplification factor and normalized sloshing force at  $x=-a$  for different middle baffle heights $(d = 0.8h,0.6h,0.4h)$ while keeping $a/h = 4.0, P= 0.2,d_2/h = 0.3, S/a= 0.3$ fixed. Increasing the baffle height reduces the resonance peak amplitude in the first, third, and fifth modes of non-dimensional frequency, as it introduces greater resistance to fluid motion, enhancing wave energy dissipation. Conversely, smaller baffle heights provide insufficient obstruction, allowing higher wave amplitudes to persist. The largest peak amplitude occurs at the lowest frequency for $d=0.6h,0.4h$ but with a narrower bandwidth compared to the third and fifth modes. However, for $d=0.8h$ the largest peak amplitude shifts to the third mode, while the bandwidth trend remains similar to the lower baffle heights.  An optimized result is achieved for $d=0.8h$, where sloshing suppression is more effective across multiple modes. Building on this, \Cref{fig:rect_tank_toppara_diff_d2} analyze the influence of extreme side baffle height $d_2$, keeping other parameter fixed $(a/h=4.0, S/a = 0.3, P=0.2, d/h =0.8)$. As the baffle height increases, the peak amplitude decreases for the first and third frequency modes, while the fifth mode exhibits the opposite trend. Additionally, the bandwidth is narrower for the first mode but progressively increases for the third and fifth modes. Hence, optimal sloshing suppression is attained when the extreme baffle height is nearly half of the middle baffle height, resulting in a well-balanced response across all frequency modes.

\subsection{Effect of spacing of baffles}
\begin{figure}[htpb]
\subfigure[]{\includegraphics[scale=0.178]{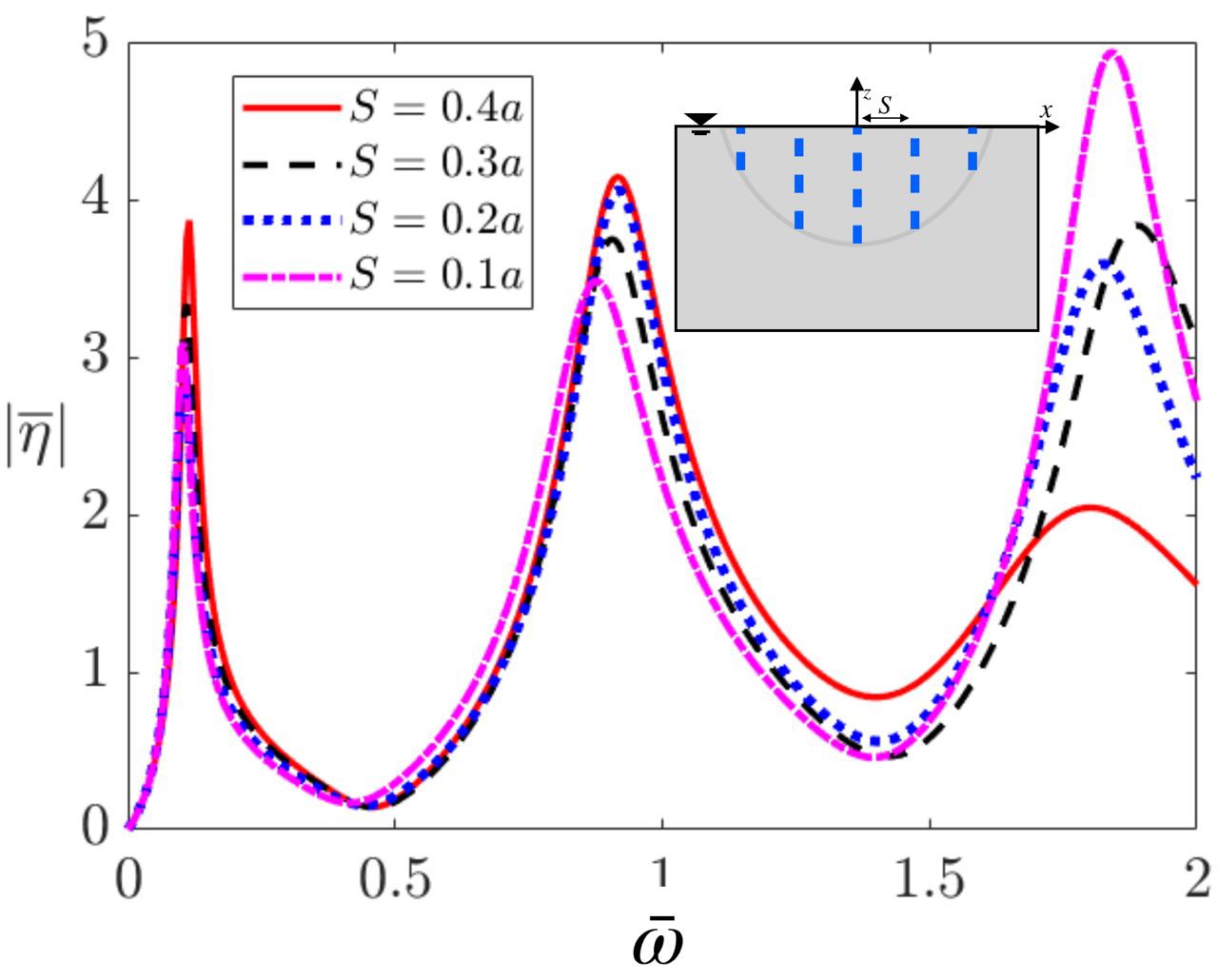}}
\subfigure[]{\includegraphics[scale=0.175]{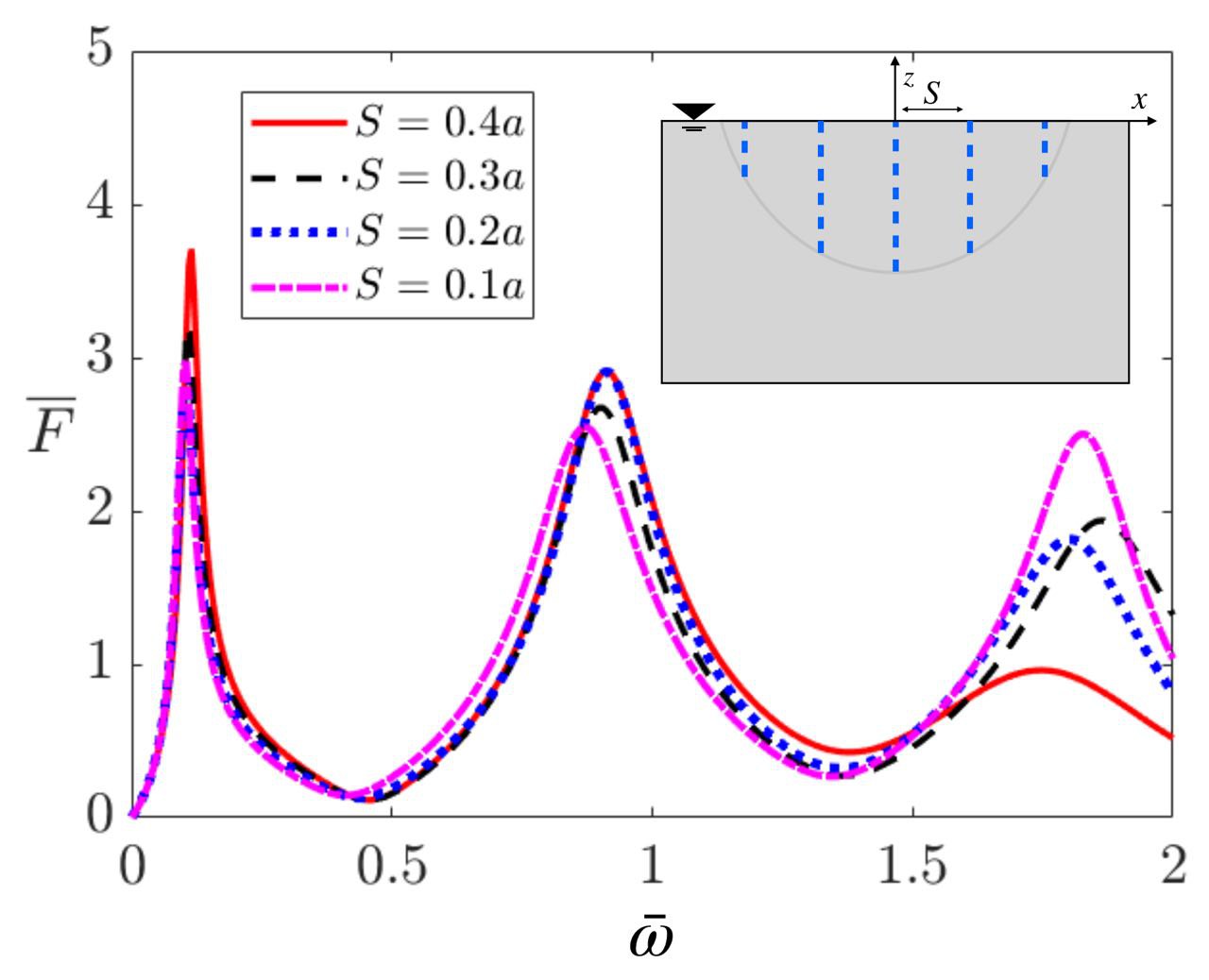}}
\caption{(a,b) Amplification factor at $x=-a$ and normalized sloshing force on tank wall for different spacing between two porous baffle $(S)$ $(a/h = 4.0, P = 0.2, d/h = 0.8, d_2/h = 0.4)$.}
\label{fig:rect_tank_toppara_diff_dx}
\end{figure}

The fluctuation of the amplification factor and normalized sloshing force is significantly influenced by the spacing between two porous baffles which is shown in ~\Cref{fig:rect_tank_toppara_diff_dx} while fixing the other physical parameter $(a/h=4.0, P=0.2, d/h=0.8, d_2/h=0.4)$. The first two resonance peaks, corresponding to the first and third frequency modes, show minimal variation across different values of 
$S$, whereas the third resonance peak exhibits a significant variation for amplication factor and normalized sloshing force. In both cases presented in \Cref{fig:rect_tank_toppara_diff_dx}, the peak heights for $S=0.3a$ and $S = 0.2a$ are similar, yet they remain distinctly separated from those corresponding to the other two $S$ values. Notably, the third resonance exhibits the lowest sloshing amplitude and force when $S=0.4a$, indicating a significant reduction in wave energy. This observation suggests that positioning the baffles closer to the sidewalls enhances their ability to suppress sloshing, likely due to increased resistance to fluid motion and improved energy dissipation.

\section{Conclusions}
In this paper, the scaled boundary finite element method has been extended to study the sloshing dynamics in rectangular tanks with porous baffles. The salient feature of the proposed framework is that only the boundary of the domain is discretized, thus reducing the computational burden without compromising on the accuracy. The robustness and the accuracy of the proposed framework is demonstrated by comparing the amplification factor and the sloshing force with results available in the literature. It is seen that the proposed framework requires at least one order fewer degrees of freedom when compared to traditional approaches. From the numerical study, it can be inferred that a fifth order polynomial yields accurate results. Different baffle arrangements in a rectangular tank have been analyzed, leading to the identification of a more effective configuration. For these configurations, optimal values of porosity, tank width-to-height ratio, submergence depth and spacing between baffles are determined to achieve maximum sloshing reduction. The numerical results presented in this study will be further validated through experimental investigations in upcoming work. Extension to quadratic boundary conditions on the baffle, sloshing dynamics in three dimensional tanks and other tank configurations are scope for future work.

\bibliographystyle{unsrtnat}
\bibliography{myRef}

\end{document}